\documentclass[%
aps,pre,amsmath,amssymb,showpacs,reprint,longbibliography,tightenlines
]{revtex4-2}
\twocolumngrid

\usepackage{lineno}
\newcommand*\patchAmsMathEnvironmentForLineno[1]{
  \expandafter\let\csname old#1\expandafter\endcsname\csname #1\endcsname
  \expandafter\let\csname oldend#1\expandafter\endcsname\csname end#1\endcsname
  \renewenvironment{#1}
     {\linenomath\csname old#1\endcsname}
     {\csname oldend#1\endcsname\endlinenomath}}
\newcommand*\patchBothAmsMathEnvironmentsForLineno[1]{
  \patchAmsMathEnvironmentForLineno{#1}
  \patchAmsMathEnvironmentForLineno{#1*}}
\AtBeginDocument{
\patchBothAmsMathEnvironmentsForLineno{equation}
\patchBothAmsMathEnvironmentsForLineno{align}
\patchBothAmsMathEnvironmentsForLineno{flalign}
\patchBothAmsMathEnvironmentsForLineno{alignat}
\patchBothAmsMathEnvironmentsForLineno{gather}
\patchBothAmsMathEnvironmentsForLineno{multline}
}
 \usepackage{xcolor}

\usepackage{mathtools}
\usepackage{color}
\usepackage{bigints}
\usepackage{multirow}
\usepackage{graphicx}
\usepackage{dcolumn}
\usepackage{bm}
\usepackage{subfigure}
\usepackage[dvipsnames]{xcolor}
\usepackage{todonotes}
\usepackage[colorlinks=true,linkcolor=purple,citecolor=Green,urlcolor=magenta]{hyperref}
\usepackage{multirow}
\usepackage{mathtools}

\newif\iffigure
\figurefalse
\figuretrue
\allowdisplaybreaks[3]
\usepackage{cprotect}

\def\SM{Supplemental Material}
\def\bv{\bar{\bm{v}}}
\def\bvx{\bar{v}_x}
\def\bvy{\bar{v}_y}

\def\dd{\mathrm{d}}

\def\rnd#1#2{\frac{\partial #1}{\partial #2}}
\def\diff#1#2{\frac{\rd #1}{\rd #2}}

\def\kB{k_{\mathrm{B}}}

\def\px{\mathtt{x}_+}
\def\py{\mathtt{y}_+}
\def\mx{\mathtt{x}_-}
\def\my{\mathtt{y}_-}
\def\dd{d_{\mathrm{p}}}
\def\xp{x_{\mathrm{p}}}
\def\yp{y_{\mathrm{p}}}

\def\bvx{\bar{v}_x}

\usepackage {cancel}

\usepackage{changes}
\def\red#1{\textcolor{black}{#1}}

\def\figsize{0.7}

\def\bx{\bm{x}}
\def\rhoe{\rho_{\mathrm{e}}}
\def\db{\lambda_{\mathrm{D}}}
\def\arctanh{\mathrm{arctanh}}
\def\arcsinh{\mathrm{arcsinh}}
\def\rd{\mathrm{d}}
\def\tpsi{\widetilde{\psi}}
\def\tz{\widetilde{z}}
\def\th{\widetilde{h}}
\def\tsigma{\widetilde{\sigma}}
\def\tchi{\widetilde{\chi}}
\def\GC{\mathrm{GC}}
\def\DH{\mathrm{DH}}
\def\psiD{\psi_{\mathrm{D}}}

\begin{document}

\preprint{\today, ver.~\number\time}

\newcommand{\titleA}{
}

\newcommand{\titleB}{
Optothermal Actuation of Unidirectional Thermo-osmotic Flows
}

\title{
\titleB
}

\author{Tetsuro Tsuji$^\dagger$}
\email{tsuji.tetsuro.7x@kyoto-u.ac.jp (corresponding author)}
\altaffiliation[Present address: ]{Department of Aeronautics and Astronautics, Kyoto University, Kyoto 615-8540, Japan}
\author{Shota Suzuki$^\dagger$}%
\author{Satoshi Taguchi$^\dagger$}%
\author{Haruya Ishida$^\ddagger$}
\author{Hideaki Teshima$^{\ddagger,\S}$}
\affiliation{%
$^\dagger$Department of Informatics, Kyoto University, Kyoto 606-8501, Japan\\
$^\ddagger$Department of Aeronautics and Astronautics, Kyushu University, Fukuoka 819-0395, Japan\\
$^\S$International Institute for Carbon-Neutral Energy Research (WPI-I2CNER), Kyushu University, Fukuoka 819-0395, Japan
}%

\date{\today}

\begin{abstract}
\noindent 
In this paper, we experimentally demonstrate the microscale direction control of thermo-osmotic flows using a focused-laser heating. The key is the off-center laser irradiation on an immobilized light-absorbing microparticle, which generates a nonuniform, asymmetric heat source. The resulting thermo-osmotic flows are evaluated using the optically trapped particle tracking velocimetry (ot-PTV), presented in our preceding paper (\href{https://doi.org/10.1103/brnj-cw44}{T.~Tsuji, et al., Physical Review Fluids 11, 034901 (2026)}). It is shown that the flow characteristics can be modulated by the ionic strength of a sample solution and/or the surface molecular coating of the substrate. In particular, the significance of ionic strength on thermo-osmotic flows \red{is} discussed based on the surface potential of the substrate measured by frequency-modulated atomic force microscopy.
\end{abstract}

\maketitle
\thispagestyle{empty}


\section{\label{sec:intro}Introduction}
\begin{figure}[bt]
    \centering
    \includegraphics[width=0.9\linewidth]{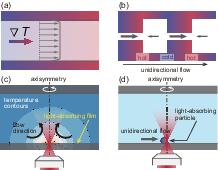}  
    \caption{(a) Thermo-osmotic flow between two parallel plates. (b) A typical Knudsen pump generating a unidirectional flow with finite temperature differences. (c) A typical optothermal-fluidic setup \cite{Tsuji2024} under thin-film heating. (d) Concept of the present experiment. Laser irradiation at the off-center position of a light-absorbing particle creates a unidirectional streamwise flow. }
    \label{fig:intro}
\end{figure}

When a solid wall has a nonisothermal temperature distribution, the adjacent fluid is driven to flow along the fluid--solid interface \red{in response to the local temperature gradient}. This flow is called thermo-osmosis (or thermo-osmotic flow) \cite{Derjaguin1987,Bregulla2016}, a type of thermally induced \red{flow}. In the case of gases, thermo-osmotic flow is known as thermal transpiration, as shown in Fig.~\ref{fig:intro}(a); see also Ref.~\cite{Sone2007} for other types of thermally induced flows. As the conversion from heat to motion, thermally induced flows are promising for micropumps without mechanically moving parts. The concept has been extensively explored as a Knudsen pump \cite{Wang2020} for gases, where unidirectional gas flows can be induced by the asymmetry of heated channels, e.g., as shown in Fig.~\ref{fig:intro}(b). As \red{a} remarkable \red{application of Knudsen-pump like structures}, optothermal levitation of tiny objects has been recently realized \cite{Cortes2020,Schafer2025,Han2025}. Note that \red{the simple} setup \red{shown} in Fig.~\ref{fig:intro}(a) requires an \red{increasingly large temperature difference as the channel length increases,}
while the periodic configuration in Fig.~\ref{fig:intro}(b) \red{requires} a finite \red{temperature difference, independent of the overall channel length.}
Herein, motivated by the success of Knudsen pumps, we consider a fluid actuation method for liquids using asymmetric heating. 

\begin{figure*}[bt]
    \centering
    \includegraphics[width=1\linewidth]{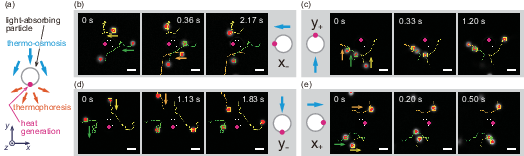}  
    \caption{Demonstration of unidirectional thermo-osmotic flow. (See also \SM~\ref{sec:SM-movie} for movie.) (a) Schematic of an experiment. A light-absorbing particle is locally heated by a focused laser propagating in the $z$ direction. A thermo-osmotic flow in the $xy$ plane is generated in the direction indicated by blue arrows. \red{The flow is investigated using tracer particles, as shown in panels (b--e). Note that thermophoresis of tracers, which is directed against the temperature gradient under the present conditions,} also occurs as indicated by orange arrows. \red{The} observed \red{tracer} motion in the experiments is the superposition of thermo-osmosis and thermophoresis. (b--e) Snapshots of tracer motions for the cases of different heating positions: (b) $\mx$ heating, (c) $\py$ heating, (d) $\my$ heating, and (e) $\px$ heating. A dashed circle at the image center \red{indicates} the light-absorbing particle. Colored arrows (yellow, orange, and green) indicate the directions of tracer motion when passing nearby the light-absorbing particle. A trapping laser is not used in these demonstrations. Scale bar: $2$~\textmu m.}
    \label{fig:exp}
\end{figure*}

Thermo-osmotic flow is affected by the molecular nature of fluids \red{and} solids and their interaction\red{s}; tuning fluid--solid intermolecular interaction\red{s} can reverse the flow direction against the temperature gradient, as demonstrated in molecular simulations \cite{Fu2017,Wang2020a,Qi2024}, microscopic theory \cite{Anzini2025}, and kinetic theory \cite{Tsuji2025}. Therefore, thermo-osmotic flow is \red{fundamentally} a molecular-scale flow. From a macroscopic viewpoint, thermo-osmotic flow manifests as \red{a} violation of \red{the} no-slip boundary condition of fluid equations due to temperature variation\red{s}. \red{Accordingly}, we introduce a thermal-slip boundary condition $\bm{v}\cdot \bm{t} = - K \nabla T\cdot \bm{t}$, where $\bm{v}$ is a flow velocity, $\bm{t}$ is a unit tangent \red{to the} boundary, $K$ is a thermal-slip coefficient, and $T$ is a local boundary temperature; this boundary condition can be used for fluid equation\red{s}, e.g. the Stokes equation\red{s}. Therefore, the thermal-slip coefficient $K$ characterizes thermo-osmotic flow. 

Let us have a quick look at the difference \red{in} $K$ between gas\red{es} and liquid\red{s}. The thermal-slip coefficient $K$ is of the order of $\nu/T\approx 10^{-7}$~m$^2$~s$^{-1}$~K$^{-1}$ for gas\red{es} \cite{Sone2007}\red{, where $\nu$ is the} kinematic viscosity \red{of} air at room temperature. For liquid\red{s}, the thermal-slip coefficient $K$ varies drastically depending on physical conditions, such as the choice of materials; $|K|$ ranges \red{from} $O(10^{-12})$ \red{to} $O(10^{-9})$~m$^2$~s$^{-1}$~K$^{-1}$ \cite{Fu2017,Ganti2017,Qi2024,Bregulla2016,Tsuji2023}\red{. These values are} significantly smaller than th\red{ose} for gas\red{es}, \red{indicating that} thermo-osmotic flow in liquid\red{s} is typically weaker than that in gas\red{es}. This weakness hinders the Knudsen-pump-like application of thermo-osmotic flow in liquid\red{s}.

To compensate \red{for} the weakness of thermo-osmotic flow, a \red{sufficiently large} temperature gradient \red{is required}. Photothermal conversion is useful here, since heat sources can be easily localized in micro- and nanoscale by focusing lasers \cite{Duhr2006} and/or utilizing the plasmonic heating of nanostructures \cite{Baffou2017}. In Ref.~\cite{Tsuji2024}, such photothermal fluid actuation, so-called optothermal fluidics, has been thoroughly reviewed and a simple semianalytical model for flow analyses has been proposed, as schematically shown in Fig.~\ref{fig:intro}(c). Usually, the laser is irradiated to microchannels (or slits) in the wall-normal direction in the majority of existing experiments. However, this setup induces axisymmetric flows about the laser axis and is not suited for the actuation of a unidirectional streamwise (or in-plane) flow. 

Considering the analogy of the gaseous Knudsen pump, the spatial symmetry of generated heat \red{needs}  to be broken to obtain unidirectional liquid flows. In the literature, multiple lasers with different powers \cite{Namura2019,Yamada2025} and engineered asymmetric nanostructures \cite{Jing2024,Tamura2022a,Setoura2025} have been proposed to break the spatial symmetry of photothermal heat sources. Here, we instead use commercially available light-absorbing particles (or thermally-active particles) \cite{Paul2022a,Rohde2025,Rohde2026}, as shown in Fig.~\ref{fig:intro}(d), because of their experimental simplicity. These microparticles are basically made of low-thermal-conductivity materials such as polystyrene \cite{Paul2022a}, SiO$_2$ \cite{Rohde2025}, and melamine \cite{Rohde2026}, containing smaller metal nanoparticles (iron oxide or gold) as heat absorber. The low-thermal conductivity \red{of these particles} enables inhomogeneous \red{heating} inside the particles upon the off-center laser irradiation (Fig.~\ref{fig:intro}(d)), which breaks the axisymmetry. In particular, Ref.~\cite{Rohde2026} proposed the generation of engineered flow fields around a light-absorbing particle toward programmable active matters, which corresponds to optothermal-levitation-like applications \cite{Cortes2020,Schafer2025,Han2025} in gases. 

In this paper, we investigate the characteristics of thermo-osmotic flow around a light-absorbing particle attached to a channel wall, i.e., a substrate, as a proof-of-concept of unidirectional thermal flow actuation. More specifically, we use the optically trapped particle tracking velocimetry (ot-PTV) \cite{DiLeonardo2006,Tsuji2026} as \red{a} pointwise flow measurement method, which is advantageous in near-wall slow microflow measurement. The effects of salt concentration (or ionic strength) \cite{Duhr2006,Vigolo2010,Eslahian2014,Syshchyk2016,Herrero2022,Ouadfel2023,Mayer2023a,Pandey2023,Chen2023b} and the surface coating \cite{Bregulla2016} of the substrate on unidirectional thermo-osmotic flow are experimentally evaluated. In addition, the physical origin of the ionic strength dependence is discussed based on the surface potential measurement using frequency-modulated atomic force microscopy (FM-AFM).

\begin{figure*}[tb]
    \centering
    \includegraphics[width=1\linewidth]{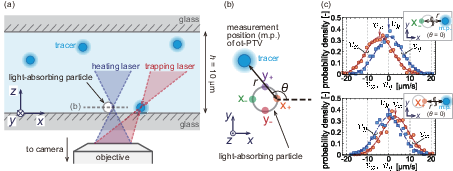}  
    \caption{(a) The ot-PTV experimental setup. Horizontal dashed line indicates the cross section observed by a camera (panel (b); see also Fig.~\ref{fig:exp}). (b) Definition of the measurement position using ot-PTV. The radial coordinate $(r,\theta)$ centered at the heating positions $\px$, $\py$, $\mx$, or $\my$ is introduced. Panel (b) corresponds to the case of the $\px$ heating. See also \SM~\ref{sec:SM-movie} for movie of ot-PTV experiment. (c) Examples of the probability distribution of the tracer velocity obtained in the ot-PTV measurement, where the insets show the schematics of the measurement positions. Symbols and curves represent the experimental results and the Gaussian fitting, respectively, for $v_x$ (orange circle) and $v_y$ (blue square). Upper (or lower) panel shows the case of the $\mx$ (or $\px$) heating with $r=5$~\textmu m and $\theta=0$. The mean of the Gaussian, $\bvx$ (or $\bvy$) is shown by arrows.}
    \label{fig:setup}
\end{figure*}


\section{\label{sec:method}Experimental Method}
We describe here an overview of our experimental setup; the details can be found in Appendix~\red{\ref{sec:SM-flow}}. 

Briefly, localized focused-laser heating of an immobilized light-absorbing particle on a substrate induces a unidirectional thermo-osmotic flow, as shown in Fig.~\ref{fig:exp}(a). The flow direction is tunable by changing the heating position within the particle in the $xy$ plane, as demonstrated in Fig.~\ref{fig:exp}(b--e) by observing tracer motions. Note that the tracer motion is the superposition of thermo-osmotic flow and thermophoresis (see Fig.~\ref{fig:exp}(a)). Panels (b), (c), (d), and (e) show the cases of (b) $\mx$ heating, (c) $\py$ heating, (d) $\my$ heating, and (e) $\px$ heating, where the symbols $\mathtt{x}_\pm$ and $\mathtt{y}_\pm$ represent that the laser focuses are placed at $(x,y)=(\xp\pm \dd/2,\yp)$ and $(x,y)=(\xp,\yp\pm \dd/2)$, respectively, with $(\xp,\yp)$ the center position of the light-absorbing particle and $\dd=1.31$~\textmu m the diameter of the light-absorbing particle. In this paper, we focus on the in-plane flow characteristics in the $xy$ direction. 

A fluidic device is a glass-made microslit of a width $h$, which is filled by a sample solution. To suppress thermal convection, the width is set to $h=10$~\textmu m \cite{Tsuji2021}. The light-absorbing particles, i.e., polystyrene microspheres containing iron oxide nanoparticles \cite{Paul2022a}, are fixed onto the surfaces of the microslit, as shown in Fig.~\ref{fig:setup}(a). The sample solution is a tris-HCl buffer (pH$=8.0$) solution of the concentration $c$ with $c=0.1$, $1$, $10$~mM and contains fluorescent polystyrene tracers of a diameter of $1$~\textmu m.

Two lasers, i.e., a heating laser with a wavelength of $488$~nm and a trapping laser with a wavelength of $1064$~nm, are irradiated to the device in the perpendicular direction to the microslit (i.e., the $z$ direction) through an oil-immersion objective (100x magnification, NA$=1.45$) of an inverted microscope, as shown in Fig.~\ref{fig:setup}(a). The laser power of the heating laser is set to $P$ with $P=7.5$, $10$, $12.5$, and $15$~mW at the laser driver, while the laser power of the trapping laser is fixed to $56$~mW after the objective. The heat generation by the trapping laser is considered negligible due to low absorption characteristics of water \cite{Ito2007} and polystyrene tracers. The heating laser locally heats up the light-absorbing particle as well as the surrounding fluid and the microslit wall. The trapping laser specifies the flow measurement position by optical trap of tracers. Figure~\ref{fig:setup}(b) is an example showing the measurement position for the $\px$ heating case, where $(r,\theta)$ defines the local radial coordinate of the measurement position with respect to the heating position. 

The repeated cycles of the trap-and-release of the tracers accumulate the velocity vectors of the tracers at the measurement position, the average of which is regarded as the ``flow" velocity $\bv=(\bvx,\bvy)$ in the $xy$ plane (ot-PTV \cite{Tsuji2026}). Note that the flow $\bv$ is considered as the superposition of thermo-osmotic advection and thermophoresis. In Fig.~\ref{fig:setup}(c), two examples of the ot-PTV results are presented for the $\mx$ heating (upper panel) and the $\px$ heating (lower panel), where $r=5$~\textmu m and $\theta=0$. In both cases, the probability distributions of $v_x$ (or $v_y$) \red{are well approximated by} Gaussian \red{distributions,} and their means are indicated by arrows. We notice from these distributions that the tracer motion toward (or away from) the heat is detected for the $\mx$ (or $\px$) heating. Considering that thermophoresis is the tracer motion repelling from the heat \red{under the present condition}, the $\mx$ heating case (i.e., the tracer motion toward the hot) drives thermo-osmotic flow around the light-absorbing particle from cold to hot. The \red{mean tracer velocity} in the $y$ direction is negligible due to the spatial symmetry, leading to $\bvy\approx 0$~\textmu m/s. 

The advantages of the ot-PTV here are \red{as follows}: 
\begin{enumerate}
    \item The $xy$ position of measurement can be set even \red{in} a tracer-sparse region created \red{by} thermophoresis pushing tracers away from the heating position (see Fig.~\ref{fig:exp}(b--e) where no tracer is observed near the heating position).
    \item \red{The $z$ position of the measurement can be \red{maintained at approximately} the same value for all runs and all tracer displacements, thereby enabling the evaluation of thermo-osmotic flow which decays \red{rapidly with increasing} distance from the wall. Here, we ensure that the $z$ position is almost the same for every run by focusing on the equatorial plane of the light-absorbing particle using a motorized focus drive (PS3H122R \& ES10ZE, Prior Scientific) with a spatial resolution of $0.1$~\textmu m. In our setup, a $0.1$~\textmu m displacement produces a noticeable change in the particle appearance under the high-NA objective (NA$= 1.45$), and thus the measurement height is considered to be reproduced within approximately $0.1$~\textmu m.} We may carry out usual particle image velocimetry (PIV) or PTV instead of ot-PTV, but in this case we cannot control the $z$ position which is expected to affect significantly the magnitude of thermo-osmotic flow.
    \item The flow analysis is possible even under experimental conditions in which tracers tend to stick to the substrate, e.g., \red{at} high salt \red
    {concentrations, where the reduced Debye length weakens electrostatic repulsion between the tracers and substrate, allowing Brownian tracers to approach the wall more easily}. 
\end{enumerate} 

\red{By analyzing the absorption characteristics of the light-absorbing particle (see Appendix~\ref{sec:abs}) at the wavelength of $488$~nm, the temperature rise in the light-absorbing particle is roughly estimated to be $O(10)$~K (see Appendix~\ref{sec:heatgeneration} and \ref{sec:temperature}).
This temperature elevation is expected to produce a temperature gradient of the order of $1$~K/\textmu m around the particle, which is sufficiently strong to induce thermo-osmotic flows. We also perform an experimental assessment of the temperature elevation. Due to the use of the visible laser for heating, we cannot use a method based on laser-induced fluorescence, which we have carried out in previous studies (e.g., Ref.~\cite{Tsuji2023}). Instead, we use a method based on the phase transition of liquid crystals \cite{Bregulla2016,Paul2022a,Rohde2026}, and the estimated temperature elevation is qualitatively supported (see Appendix~\ref{sec:5CB}).}


\section{Results}\label{sec:result}
\subsection{Competition between thermo-osmosis and thermophoresis} 
\begin{figure}[tb]
    \centering
    \includegraphics[width=\linewidth]{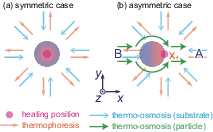}  
    \caption{Schematics of the relation between the heating position and the resulting thermally induced flows. (a) Symmetric case with heating at the center. (b) Asymmetric case with heating at the off-center position $\px$. }
    \label{fig:contribution}
\end{figure}

Figure~\ref{fig:contribution} schematically shows possible contributions to the tracer motion driven by local heating, namely, thermophoresis and thermo-osmosis. Here, thermo-osmosis is driven over (i) the substrate and (ii) the light-absorbing particle. Note that, thermal convection is negligible in our setup due to a tight confinement (i.e., $h=10$~\textmu m). The effect of the optical force of heating laser is also negligible since measurement positions are set to $r=5$~\textmu m (see Fig.~\ref{fig:setup}(b)), which is sufficiently far from the focus of the heating laser. \red{This position $r=5$~\textmu m is chosen because the tracer motion is found to be most significant in a preliminary PIV measurement.}

First, let us look at the symmetric case, where the heating laser is focused at the center of the light-absorbing particle (Fig.~\ref{fig:contribution}(a)). This case was already demonstrated in our previous work \cite{Tsuji2026}, Fig.~8 there. Recall that thermophoresis of the tracer is directed from hot to cold (orange arrows in Fig.~\ref{fig:contribution}) while the thermo-osmotic flow over the substrate  (cyan arrows in Fig.~\ref{fig:contribution}) is opposite. Then, the tracers tend to stay at a specific radial position at which the thermophoretic force and the drag force by thermo-osmotic flow counterbalance. 

Next, let us consider the asymmetric case, where the heating laser is focused at the off-center position $\px$. Now, in addition to the effects of thermophoresis and the thermo-osmotic flow over the substrate, there arises the directed thermo-osmosis over the light-absorbing particle, schematically illustrated by green arrows in Fig.~\ref{fig:contribution}(b). Moreover, the temperature profile is not axisymmetric due to the thermal-conductivity mismatch between the particle and the fluid. As a result, the tracer at the position A (or the position B) is repelled from (or pulled to) the heating position $\px$, producing a unidirectional motion. 
However, temperature imaging of such asymmetric state in microscale is difficult for our present experimental setup; Ref.~\cite{Rohde2026} presented a temperature measurement using a liquid crystal for a similar experimental condition, confirming asymmetric temperature profile. 

\begin{figure}[tb]
    \centering
    \includegraphics[width=\figsize\linewidth]{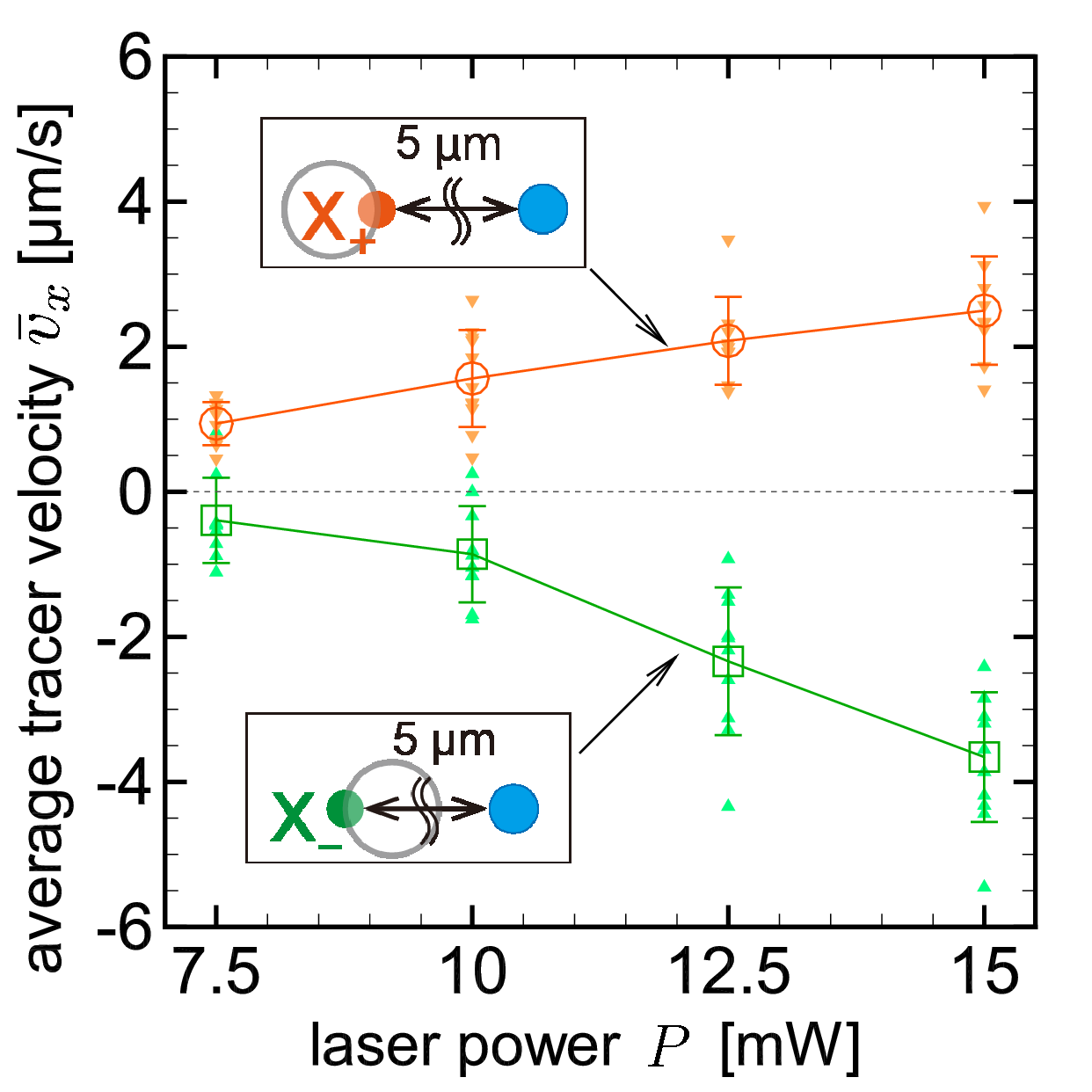}  
    \caption{Typical ot-PTV results on the average tracer velocity $\bvx$ for various heating laser power $P$ with $r=5$~\textmu m and $\theta =0$ (cf. Fig.~\ref{fig:setup}(b)). The heating position is set to $\px$ (orange circle) or $\mx$ (green square). The ionic concentration is $c=1$~mM.}
    \label{fig:11}
\end{figure}

\begin{figure}[tb]
    \centering
    \includegraphics[width=\figsize\linewidth]{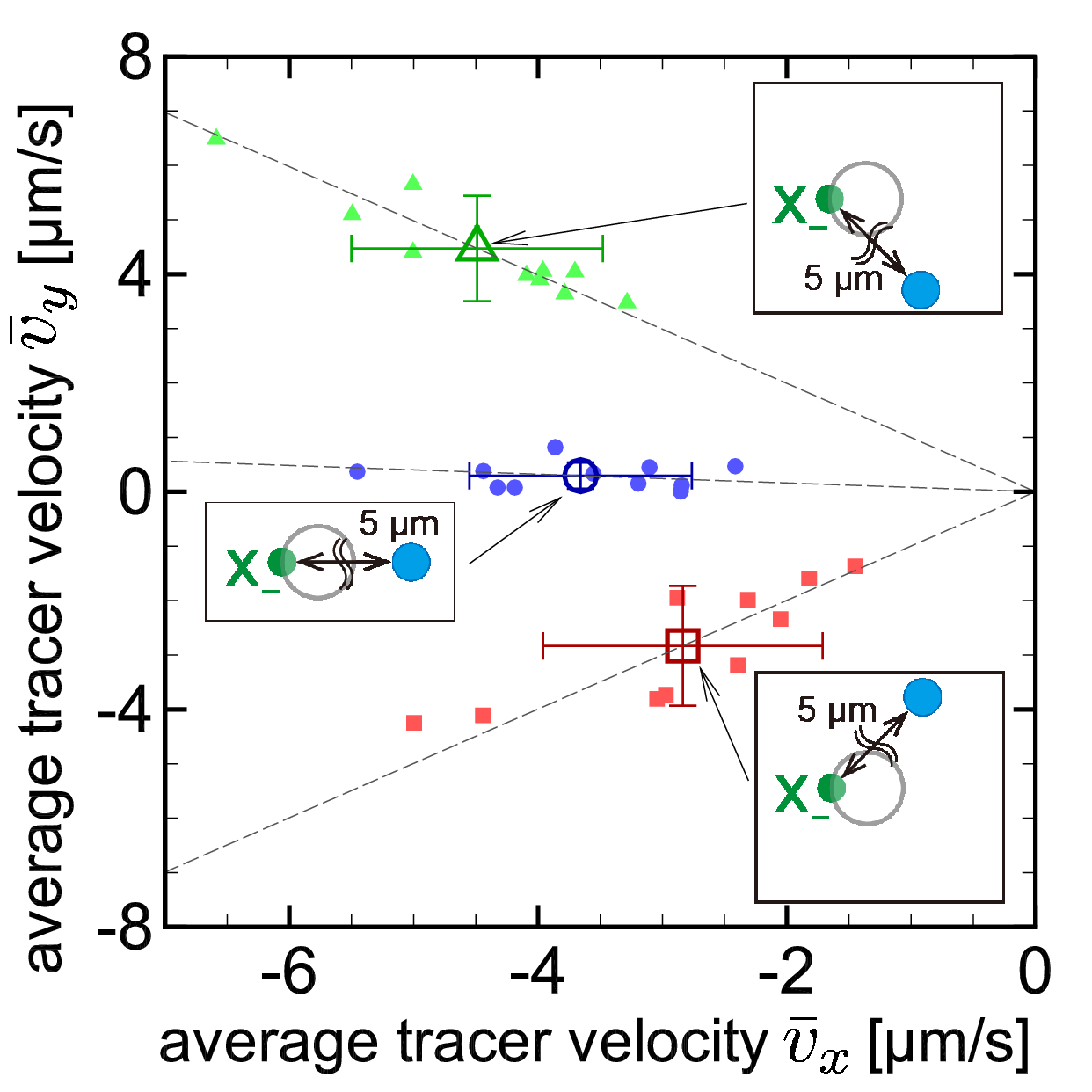}  
    \caption{Tracer velocity vector $(\bvx,\bvy)$ for $P=15$~mW with different measurement positions: $\theta= 0$ (blue circle) $\theta=\pi/4$ (red square), and $\theta=-\pi/4$ (green triangle) with $r=5$~\textmu m. Dashed lines connect the origin and the average points. The ionic concentration is $c=1$~mM.} 
    \label{fig:10}
\end{figure}

Figure~\ref{fig:11} shows typical results of $\bvx$ obtained by ot-PTV for various heating laser \red{powers} $P$ with the ionic concentration $c=1$~mM. During the experiments, we \red{observe} moderate \red{variations in} $\bvx$ \red{among} individual light-absorbing particles. This may be \red{attributed} to differences \red{among individual} light-absorbing particles \red{(see Appendix~\ref{sec:particle}),} the nonuniform distribution of \red{iron-oxide} nanoparticles within the light-absorbing particles\red{, and/or errors in the positioning of the heating laser irradiation ($\leq 100$~nm)}. Therefore, we \red{repeat the} measurements about ten times for \red{each} experimental condition, replacing the light-absorbing particle and the tracer \red{with} new ones before each run. In Fig.~\ref{fig:11}, small symbols show the results of \red{individual runs} and large symbols their averages. 

In the case of the $\px$ heating (orange circle), Fig.~\ref{fig:11} shows that the tracer is repelled from the heat when it is placed in the same side of heating, corresponding to the position A in Fig.~\ref{fig:contribution}(b). The speed of depletion increases with $P$ while $\bvx$ remains below $3$~\textmu m/s. Note that the further increase of the laser power often results in the creation of bubbles, which significantly disturb the experiments. Conversely, in the case of the $\mx$ heating (green square), the tracer is attracted to the heat when it is placed in the opposite side of heating, corresponding to the position B in Fig.~\ref{fig:contribution}(b). 
In the absence of thermal convection and optical trapping (of the heating laser), the directed motion toward heat is thermo-osmotic flow. 
 
To investigate the spatial characteristics of thermo-osmosis, Fig.~\ref{fig:10} shows the results of the tracer velocity $(\bvx,\,\bvy)$ at three different measurement positions. We set $r=5$~\textmu m with the $\mx$ heating, while $\theta$ is set to $\pm \pi/4$ and $0$ (see the insets of Fig.~\ref{fig:10}). For $\theta=0$ (blue-circle symbols), as already described in Fig.~\ref{fig:11}, the tracer is directed toward the heat, with the $y$ component being nearly zero. For $\theta= \pi/4$ (red-square symbols) and $-\pi/4$ (green-triangle symbols), the plots are distributed on the lines $\bvy\approx \bvx$ and $\bvy\approx -\bvx$, respectively, indicating the tracer motion toward the heat. These motions are consistent with the observation in Fig.~\ref{fig:exp} and movie~\ref{sec:SM-movie} without optical trapping\red{.}

\subsection{Enhancing thermo-osmotic flow by tuning ionic strength or surface coating} 

\begin{figure}[tb]
    \centering
    \includegraphics[width=\figsize\linewidth]{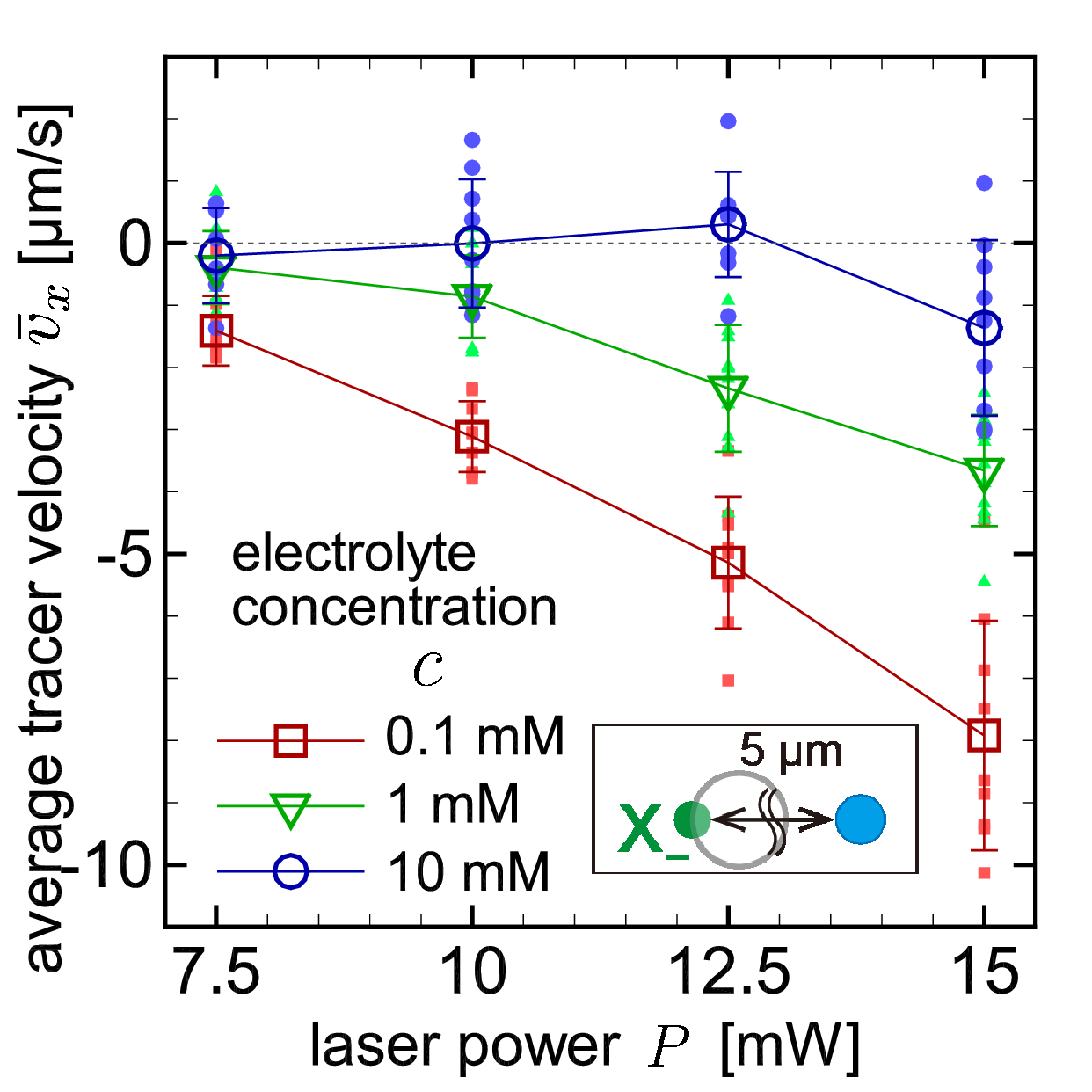}  
    \caption{Salt dependence analysis of $\bvx$ for various heating laser power $P$. The results of ot-PTV for the $\mx$ heating are summarized, where the ion concentrations are set as $c=0.1$~mM (red square), $1$~mM (green lower triangle), and $10$~mM (blue circle). }
    \label{fig:12}
\end{figure}

Thermo-osmotic flow is considered to depend on electrical properties of \red{the} fluid and \red{the} solid \red{surface}, such as ionic strength and surface charge density (e.g., Refs.~\cite{Fayolle2008,Herrero2022}). Figure~\ref{fig:12} shows the ionic concentration dependence of the thermo-osmotic flow in the $\mx$ heating case. To be more precise, in addition to the case of $c=1$~mM in Fig.~\ref{fig:11}, we show the cases of $c=0.1$~mM and $10$~mM in Fig.~\ref{fig:12}. At the highest heating laser power $P=15$~mW, \red{lower} (or \red{higher}) ionic strength enhances (or reduces) the magnitude of $\bvx$, \red{indicating a change in} the thermo-osmosis contribution. For instance, the magnitude $|\bvx|$ for $c=0.1$~mM is $2.9\pm0.7$ times that for $c=1$~mM. \red{At lower laser powers} ($P=7.5$--$12.5$~mW), the results with $c=10$~mM show almost zero velocity, possibly due to the cancellation between thermophoresis and thermo-osmotic flow. \red{We remark here that the contribution of tracer thermophoresis, which is directed from hot to cold under the present conditions, is considered to be more significant at lower $c$ (see Appendix~\ref{sec:thermophoresis}). Thus, the enhanced tracer motion toward the hotter region at lower $c$, observed in Fig.~\ref{fig:12}, cannot be explained by tracer thermophoresis alone, but instead indicates an enhancement of the thermo-osmotic flow.}

\begin{figure}[tb]
    \centering
    \includegraphics[width=\figsize\linewidth]{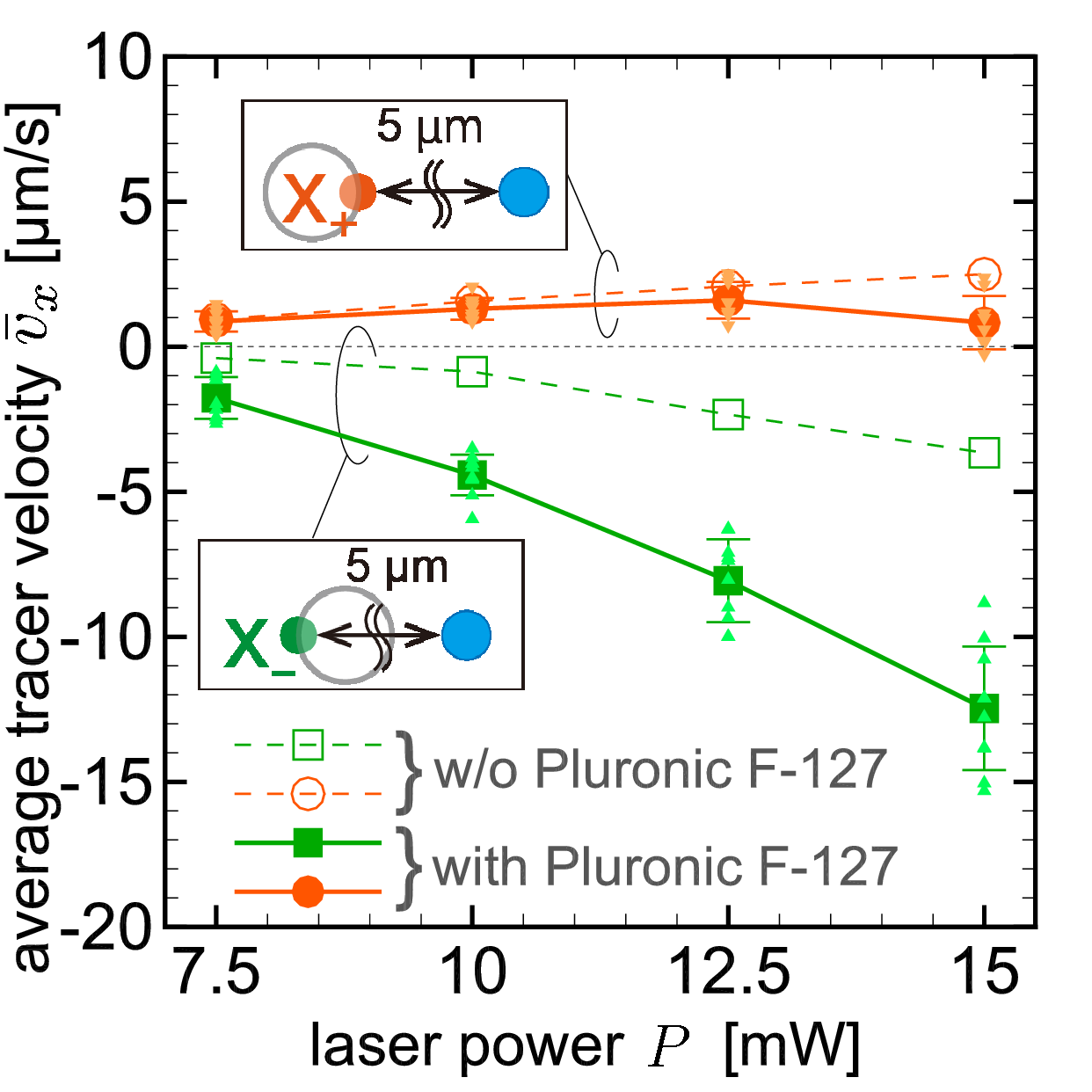}  
    \caption{Effect of the surface coating of the substrate by a nonionic surfactant Pluronic F-127. The average velocities $\bvx$ for various heating laser power $P$ and for the $\mx$ and $\px$ heating are summarized, where filled and empty symbols represent the cases with Pluronic coating and without coating, respectively. }
    \label{fig:add}
\end{figure}

Since thermo-osmotic flow is driven at the fluid--solid interface, its characteristics should depend on the molecular-level properties of the interface. It \red{has been reported} that, when the surface of the substrate is covered by a nonionic block copolymer Pluronic F-127, thermo-osmotic flow is significantly enhanced \cite{Bregulla2016}.
We demonstrate the same trend here, and the results are summarized in Fig.~\ref{fig:add}, where the velocity $\bvx$ for the $\px$ and $\mx$ heating is presented. For comparison, the cases without and with coating are both shown by \red{open} and filled symbols, respectively, \red{at} $c=1$~mM. In the case of $\mx$ heating, the magnitude of $\bvx$ \red{with coating} is $4.1\pm0.7$ times \red{that without coating}. \red{We consider that the nonionic polymer mainly modifies the substrate surface, and thus the observed enhancement may be attributed to a change in the thermo-osmotic flow along the substrate.} In Ref.~\cite{Bregulla2016}, the thermo-osmotic coefficients were $1.8\times10^{-10}$~m$^2$/s and $13\times10^{-10}$~m$^2$/s for the cases without coating and with coating, respectively, \red{corresponding to a} $7.2$-fold increase. Here, we obtain \red{a} similar but smaller enhancement. In the case of $\px$ heating, thermophoresis (positive $\bvx$ contribution) is counterbalanced by enhanced thermo-osmotic flow (negative $\bvx$ contribution) at $P=15$~mW, \red{resulting in a} decreased magnitude \red{of $\bvx$}. 

\def\pa{\phantom{1}}
\begin{table*}[bt]
\renewcommand{\arraystretch}{1.5}
    \centering
    \setlength{\tabcolsep}{1em}
    \caption{The \red{magnitude of the diffuse-layer potential $|\psiD|$} and \red{Debye length} $\db$ \red{obtained from} FM-AFM \red{measurements, together with the estimated surface charge density $|\sigma|$.} \red{The surface potential $\psi_0$ and Debye length $\db$ calculated using} \red{the} Grahame \red{equation}, Eq.~\eqref{eq:table}\red{, are also shown for comparison}.}
    \label{tab:AFM}
    \begin{tabular}{ccccccc}
    \hline
    &\multicolumn{3}{c}{FM-AFM experiment}& \quad &\multicolumn{2}{c}{Eq.~\eqref{eq:table} with $\sigma=-0.32$~mC/m$^2$ \cite{Behrens2001}}\\
    \cline{2-4}\cline{6-7}
        $c$ [mM] & $\red{|\psiD|}$ [mV] & $\db$ [nm] & $|\sigma|$ [mC/m$^2$] & & $\psi_0$ [mV] & $\db$ [nm] \\
        \hline
        $0.1$& $9.0\pm1.3$&    $12.5\pm1.8$ &$0.49\pm0.02$ && \red{$-18.4$}&\red{$41.6$}\pa \\
        $1  $& $8.1\pm0.8$& \pa$8.9\pm1.0$  &$0.62\pm0.04$ && \pa\red{$-6.0$}& \red{$13.1$}\pa\\
        $10 $& $5.6\pm0.7$& \pa$6.2\pm1.3$  &$0.63\pm0.09$ && \pa\red{$-1.9$}& \red{$4.1$}\\
        \hline
    \end{tabular}
\end{table*}

\section{Discussion}
Due to the complex interplay among thermo-osmotic flow over the substrate, that over the light-absorbing particle, and thermophoresis, understanding the observed tracer motion is challenging. Here, as a first step, we try to examine the \red{the effect of} ionic concentration $c$, based on \red{a} fluid-dynamic approach \red{incorporating} electrostatic effects \cite{Fayolle2008}. 

We express the thermal-slip coefficient $K$ as $K = \red{-} \chi/T$, where $\chi$ is called a thermo-osmotic coefficient \cite{Bregulla2016}. Under some reasonable assumptions, we obtain 
\begin{align}
\chi=\varepsilon \psi_0^2 / 8 \eta, \label{eq:chi}
\end{align}
where $\varepsilon$ is the dielectric permittivity of water, $\psi_0$ is the surface electric potential at the liquid--solid interface, $\eta$ is the viscosity \cite{Bregulla2016,Fraenzl2022}; see \SM~\ref{sec:SI-theory} for details. \red{At the} relatively \red{low} ionic \red{concentration} investigated in this paper, $c\leq 10$~mM, the dielectric permittivity of water can be considered constant with respect to $c$ \cite{Gavish2016}. Therefore, we expect that the trend in Fig.~\ref{fig:12}, where the thermo-osmotic flow toward the heat source (i.e., negative velocity) is enhanced for smaller $c$, \red{may} be explained \red{by} the dependence of $\psi_0$ on the ionic concentration $c$.

\red{The} surface potential $\psi_0$, surface charge density $\sigma$, and bulk ionic number density $n_0(=\red{I_{\rm ion}} N_A)$ (\red{$I_{\rm ion}$ is the ionic strength in $\mathrm{mol}/\mathrm{m}^3$ and} $N_A$ is the Avogadro \red{constant}) are related as
\begin{subequations}\label{eq:table}
\begin{align}
    &\sinh(\frac{e\psi_0}{2\kB T})=\frac{\sigma}{\sqrt{8\varepsilon n_0 \kB T}} \left(= \frac{\sigma \db e}{2\varepsilon \kB T}\right), \label{eq:grahame} \\
    &\db = \sqrt{\frac{\varepsilon \kB T}{2 e^2 n_0}}, \label{eq:db}
\end{align}
\end{subequations}
where $\kB$ \red{is} the Boltzmann constant, $e$ is the elementary charge, and $\db$ is the Debye length; Eq.~\eqref{eq:grahame} is \red{known as} \red{the} Grahame \red{equation}. 
\red{Although the direct measurement of $\psi_0$ is difficult, it has been reported that $\psi_0$ is well correlated with the zeta potential \cite{Brown2016} as the ionic concentration $c$ varies. Under the assumption that the zeta potential can be approximated by the diffuse-layer potential $\psiD$, we characterize $\psiD$ experimentally instead of $\psi_0$ in the following.}

Using frequency-modulation atomic force microscopy (FM-AFM) (see Appendix~\ref{sec:SM-AFM}), we experimentally evaluate the \red{diffuse-layer potential $\psiD$} and the Debye length $\db$ at the interface between the bottom glass substrate and the electrolyte solution for $c=0.1$, $1$, and $10$~mM, i.e., the same conditions used in Fig.~\ref{fig:12}. The results are summarized in Table~\ref{tab:AFM}, where we show \red{only} the magnitude of \red{$\psiD$} because the FM-AFM approach cannot determine the sign. The \red{magnitude of the} surface charge density\red{, $|\sigma|$,} is \red{then estimated} from Eq.~\eqref{eq:grahame} using \red{the measured $|\psiD|$} and $\db$\red{, with $\psiD$ used as an approximation to $\psi_0$}. \red{Note that, because $\psi_0$ and $\psiD$ are generally different, the estimated surface charge densities should be regarded as approximate values.}

In the table, \red{$|\psiD|$} and $\db$ show \red{a} decreasing trend as $c$ increases. This trend is qualitatively consistent with \red{the trend observed in} Fig.~\ref{fig:12} and \red{the prediction based on} Eq.~\eqref{eq:chi}, i.e., thermo-osmotic flow velocity (or thermo-osmotic coefficient $\chi$) decreases as $c$ increases. This consistency supports the idea that thermo-osmotic flows are enhanced \red{by changing} the electrolyte concentration. To further validate our consideration based \red{on} Eqs.~\eqref{eq:chi} and \eqref{eq:table}, we use the typical literature value of $\sigma=-0.32$~mC/m$^2$ for glass \cite{Behrens2001} and the experimental condition \red{$n_0=I_{\rm ion}N_A$, where $I_{\rm ion}$ is the ionic strength computed for $c=0.1$, $1$, $10$~mM, to estimate} the surface potential $\psi_0$ and the Debye length $\db$\red{. The results are shown in Table~\ref{tab:AFM}, using Eq.~\eqref{eq:table} under a constant-charge condition.} \red{(Note that, for a given ionic strength, the assumed value of $\sigma$ affects only the calculated $\psi_0$, not $\db$.)} Although \red{the agreement is} qualitative, the FM-AFM \red{measurements and the theoretical estimates based on Eq.~\eqref{eq:table} and the literature values} in Table~\ref{tab:AFM} \red{show the same trend with increasing ionic concentration}, supporting the idea that the ionic concentration modulates the surface \red{and diffuse-layer potentials} and thus thermo-osmotic flows. \red{It should be noted that the discrepancy between the FM-AFM measurements and the theoretical estimates based on Eq.~\eqref{eq:table} is larger for smaller $c$. This may be attributed to the increased relative influence of ionic impurities and and changes in ionic concentration due to evaporation of the sample solution at lower $c$.}



\section{Conclusion}\label{sec:conclusion}

In this paper, we propose \red{a method for} optical-heating-based flow actuation and control at \red{the} microscale. \red{Off}-center heating of a light-absorbing particle, \red{which breaks} the spatial symmetry, produces \red{an asymmetric} temperature field that drives unidirectional thermo-osmotic flow. In addition to \red{demonstrating control over the flow direction}, we carry out ot-PTV \red{analysis to characterize the flow}, revealing that \red{factors} such as electrolyte concentration \red{and} molecular surface coating can affect thermo-osmotic flows. 
\red{FM-AFM measurements of the diffuse-layer potential further suggest that the enhancement of thermo-osmotic flow at lower electrolyte concentrations is associated
with an increase in the magnitude of the surface potential.}

\appendix

\section{MATERIALS AND METHODS}\label{sec:SM-method}

\subsection{Flow analysis using ot-PTV}\label{sec:SM-flow}

\begin{figure*}[bt]
    \centering
    \includegraphics[width=0.9\linewidth]{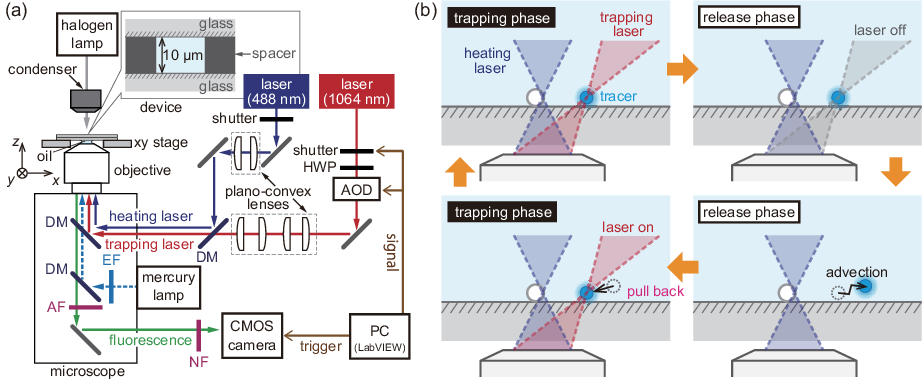}
    \caption{(a) Schematic of the experimental setup. AF: absorption filter, DM: dichroic mirror, EF: excitation filter, AOD: acousto-optic deflector, NF: notch filter for blocking $488$~nm (NF488-15, \red{T}horlabs). (b) Schematic of ot-PTV. \red{While} the trapping laser is off, the tracer \red{moves under the influence of} the local flow. After the data acquisition, the trapping laser is turned on to pull the tracer \red{back} to \red{its} initial position. The \red{on/off} switching of the trapping laser is \red{achieved} by \red{shifting} the laser focus using \red{the} AOD.}
    \label{fig:SM-optical-setup}
\end{figure*}

\noindent \textit{Optical setup}:
A trapping laser (wavelength $1064\pm 1$~nm; PowerWave 1064, NPI Laser) and 
a heating laser (wavelength $488\pm5$~nm; 06-MLD, Cobolt) are both guided to an inverted microscope (IX-73, Olympus), as shown in Fig.~\ref{fig:SM-optical-setup}(a). The microscope \red{is equipped with} two dichroic mirror units: one for the lasers and the other for fluorescent imaging. 
An objective lens (numerical aperture $=1.45$; UPLXAPO100XO, Olympus) is used to illuminate a fluidic device on \red{a motorized} $xy$ stage \red{(BIOS-L101T-S/AO/J, Olympus, resolution $0.1$~\textmu m)} and to observe the fluorescence from tracers dispersed in the device using a CMOS camera (Zyla 5.5, Andor Technology). 
\red{The fluorescence excitation} light source is a mercury lamp (U-HGLGPS, Olympus) with a fl\red{u}orescence filter unit (U-FUW, Olympus)\red{,} which is designed for excitation at wavelength\red{s} of $340$--$390$~nm and emission at wavelength\red{s} \red{above} $420$~nm. For ot-PTV, an acousto-optic deflector (AOD; DTD-276HDM \& DE-272M, IntraAction) is used to switch rapidly between trap\red{ping} and release phases. Absorption filters are \red{placed in front of} the camera to remove scattered \red{laser light}.

\vspace{0.5em}
\noindent \textit{Sample preparation}: \red{An aqueous} buffer solution contains tris-HCl (pH $=8.0$; Nippon Gene) with a concentration of $c$ ($=0.1$--$10$~mM) to \red{maintain the} pH \red{during the} experiments and to \red{vary the} ionic strength. In the solution, fluorescent polystyrene tracers with a diameter of $1$~\textmu m (F8815, Thermo Fisher Scientific) are dispersed at \red{a} concentration of $1\times10^{-2}$~wt\%. The absorption \red{peak} of the tracers \red{is at approximately} $350$~nm, well separated from the wavelength of the heating laser, $488$~nm, \red{thereby minimizing} undesired \red{changes in fluorescence} intensity \red{caused by} heating laser irradiation. 

\vspace{0.5em}
\noindent \textit{Device preparation}: Two cover glass substrates (thickness $=0.15$~mm, C022221, Matsunami Glass) are immersed in ethanol and then sonicated at $40$~kHz for $5$ minutes using an ultrasonic cleaner (MCS-6, As One Corporation). After rinsing with fresh ethanol, the substrates are \red{blown dry with compressed air}. Furthermore, the substrates are processed by atmospheric plasma exposure (BD20ACV, Electro-Technic Products Inc.) for $90$ s to eliminate residues and make the surface hydrophilic. The dispersion of light-absorbing particles ($5$~w\red{t}\%, \red{Lot: }PS-MAG-AR1065, microParticles GmbH) with a diameter of $1.31\red{\pm0.06}$~\textmu m is diluted 4000\red{-fold} using ethanol and dripped onto the cleaned substrates. \red{These particles are made of polystyrene and contain superparamagnetic iron oxide, Fe$_2$O$_3$/Fe$_3$O$_4$, at a weight fraction of $>30\%$.} \red{Further} details of the light-absorbing particles, such as the scanning electron microscope images and the absorption \red{spectrum}, can be found in Ref.~\cite{Paul2022a}. \red{After allowing the ethanol to evaporate} for $90$~minutes, the light-absorbing particles \red{become immobilized on} the \red{substrate} surfaces. Two pieces of double-sided tape with a thickness of $10$~\textmu m (NEOFIX10, NEION Film Coatings Corp.) are pasted on one of the substrates in parallel. After \red{carefully placing} the other substrate on the tapes to \red{form} a microslit device, the sample solution containing the tracers is \red{introduced into the gap} between the two substrates. To \red{prevent} drying and undesired background flows, a coverslip sealant (CoverGrip, Biotium) is \red{applied} to the open sides of the device. Before starting the experiments, we \red{verify that} the \red{gap} height is \red{approximately} $10$~\textmu m using the microscope. 

\vspace{0.5em}
\noindent \textit{Surface coating}:
\red{For experiments with} surface-coated \red{substrates}, we use a nonionic surfactant\red{,} Pluronic\textsuperscript{\textregistered} F-127, $10$~\% in H$_2$O (Biotium). \red{A} diluted Pluronic solution ($1$~\%) is \red{applied to} the cleaned surface\red{s} of the glass substrates after plasma exposure as \red{described} above. \red{The substrates are left} for $10$~minutes. The substrates are \red{gently rinsed with} ultrapure water and  \red{blown dry with compressed air}. \red{T}he dispersion of light-absorbing particles \red{is then deposited onto the substrates}. \red{The} subsequent \red{steps} are the same as \red{those} described above \red{for uncoated substrates}. This \red{coating} is expected to \red{enhance} thermo-osmotic flow \cite{Bregulla2016}.

\vspace{0.5em}
\noindent \textit{Setup of ot-PTV}:
The ot-PTV measurement \cite{Tsuji2026} consists of repeated cycles of a trap\red{ping} phase and a release phase, as shown in Fig.~\ref{fig:SM-optical-setup}(b). 
In the trap\red{ping} phase, a tracer is optically trapped at a measurement position, and in the release phase, the tracer \red{moves under the influence of} fluid flow. \red{In} our experiments, the trap duration $T_\text{trap}=60$~ms and the release duration $T_\text{release}=40$~ms are used, that is, the cycle rate is $10$~Hz. In a single cycle, we acquire three snapshots of tracer positions \red{at} a rate of $80$~Hz ($12.5$~ms \red{between} snapshot\red{s}) only in the release phase, yield\red{ing} two tracer \red{velocity} vectors \red{during} post\red{-}process\red{ing} using ImageJ with \red{the} PTA (Particle Track and Analysis ver 1.2) plug-in. For a single measurement, the number of samples acquired is $N=1000$ \red{over a} measurement time \red{of} $50$~s.  

\vspace{0.5em}
\noindent \textit{Procedure of ot-PTV}:

\noindent (1) The device is set on the \red{motorized} $xy$ stage of the microscope. We check that the height of the slit is about $10$~\textmu m using the inverted microscope and \red{there is} no leak\red{age}.  
    
\noindent (2) When the heating laser is focused on the light-absorbing particle, a bright point is observed, as shown in the raw movie data provided in \SM~\ref{sec:SM-movie}, possibly due to \red{optical scattering by the iron-oxide nanoparticles}. This bright point is \red{regarded} as the heating position. 
    
\noindent (3) A single tracer is optically trapped by a trapping laser. Using \red{the} $xy$ stage, the trapped tracer \red{is moved} to a measurement position. 

\noindent (4) \red{The heating laser is focused at one of the points} $\px$, $\mx$, $\py$, or $\my$ (Figs.~\ref{fig:exp}(b--e) and \ref{fig:setup}(b))\red{,} and ot-PTV measurement is initiated. After a single run of the experiment, the light-absorbing particle and the tracer are replaced by new ones to avoid uncontrollable degradation of the light-absorbing particle due to laser irradiation. 

\red{We comment on the precision of the heating laser position. The minimum displacement of the motorized $xy$ stage is $0.1$~\textmu m, which limits the resolution of the irradiation position.
We first set the laser position as close as possible to the particle rim with a positional resolution of $0.1$~\textmu m. Then, we move the heating position by $0.1$~\textmu m toward the particle center to ensure that the laser irradiates the particle. At present, it is difficult to quantify how the uncertainty in the irradiation position affects the thermo-osmotic flow because of the limited resolution of our experimental apparatus.}

\subsection{Surface potential analysis using AFM}\label{sec:SM-AFM}
Surface potentials of the substrates in tris-HCl aqueous solutions are estimated by frequency-modulation atomic force microscopy (FM-AFM). The measurements are performed using an AFM system (SPM-8100FM, Shimadzu Corp.) equipped with a spherical AFM probe (Biosphere B300-NCH, Nanotools). \red{The} FM mode provides higher force sensitivity than typical contact mode \cite{Ishida2026}. The AFM was operated in ZX scan mode, and the frequency shift $\Delta f$ is measured as a function of the tip-sample distance $h_d$, i.e., $\Delta f = \Delta f(h_d)$. The probe radius used in the analysis was $R_{\mathrm{probe}} = 300$~nm. 

The measurements are conducted in tris-HCl solutions with concentrations of $0.1$, $1$, and $10$ mM. During the measurement, the tip approaches the substrate surface, and the distance-dependent frequency shift $\Delta f(h_d)$ is obtained. The resonance frequency and spring constant of the cantilever are $f_0$ = $136.6$~kHz and $k_{\mathrm{spring}}$ = $24.4$~N/m, respectively.

The electric double-layer interaction is analyzed using a charge-regulation model based on the Poisson-Boltzmann theory \cite{Trefalt2016}. The regulation parameter is fixed at $p = 0.5$, corresponding to an intermediate condition between the constant-potential and constant-charge limits. The frequency shift $\Delta f(h_d)$ due to the electric double-layer interaction is fitted using
\begin{align}
\Delta f = \frac{2 \pi f_0 \varepsilon \kappa^2 \psiD^2 R_{\mathrm{probe}}}{k_{\mathrm{spring}}}
\frac{\exp(-\kappa h_d)}{[1+(1-2p)\exp(-\kappa h_d)]^2}, \label{eq:AFM}
\end{align}
where $\kappa=\db^{-1}$ is the inverse Debye length (see also \SM~\ref{sec:SI-theory}), and $\psiD$ is the diffuse-layer potential. In the fitting procedure, $\psiD$ and $\kappa$ are treated as fitting parameters. In the present paper, \red{$\psiD$ is used as a
qualitative indicator of the surface potential $\psi_0$}. 

\begin{figure*}[bt]
    \centering
    \includegraphics[width=0.9\linewidth]{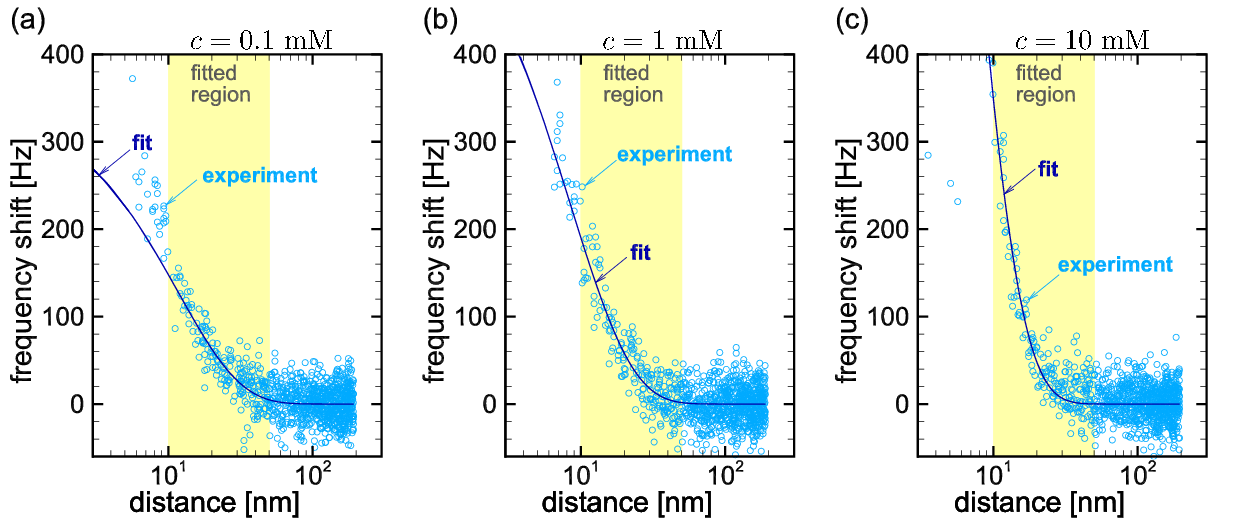}
    \caption{Frequency shift $\Delta f$ as a function of the distance $h_d$ between the probe and the substrate for (a) $c=0.1$~mM, (b) $c=1$~mM, (c) $c=10$~mM. Symbols: experimental data obtained using FM-AFM. Solid curves: curve fit using Eq.~\eqref{eq:AFM}\red{.}}
    \label{fig:SM-AFM}
\end{figure*}

Figure~\ref{fig:SM-AFM} shows the experimental results of $\Delta f(h_d)$ for (a) $c=0.1$~mM, (b) $1$~mM, and (c) $10$~mM. \red{The experimental data are reasonably well fitted by Eq.~\eqref{eq:AFM},} except \red{in the} near-wall region $h_d<10$~nm, where the \red{interaction} between the probe and the substrate may \red{become significant}.

\red{We briefly comment on the choice of $p=0.5$. We adopt this value as a reference, noting that $p=0.52$ was reported for a silica surface in Ref.~\cite{Rentsch2006}. Furthermore, we confirm that varying $p$ from $0.5$ to $0.4$ or $0.6$ changes $\psiD$ by at most $2.3\%$ and $\db$ by at most $9.6\%$, without altering the qualitative trends. The sensitivity to $p$ is stronger at lower $c$. A more detailed investigation of the effect of $p$ is needed for a quantitative analysis and is left for future work.}
        
\section{\red{Qualitative estimate of generated heat and temperature elevation}\label{sec:estimate}}
\subsection{\red{Absorption characteristics of materials\label{sec:abs}}}
\red{We compare the orders of magnitude of the absorption coefficients of water, glass (the channel walls), and the light-absorbing particle at a wavelength of $488$~nm to show that optical heating mainly occurs within the particle.}

\red{
The absorption coefficients of water and soda-lime glass at a wavelength of $488$~nm are $\alpha_{\mathrm{water}}=1.7\times10^{-2}$~m$^{-1}$ \cite{Hass1977} and $\alpha_{\mathrm{glass}}=4.3$~m$^{-1}$ \cite{Rubin1985,Polyanskiy2024}, respectively. The glass used in this study is borosilicate glass, but its absorption coefficient is expected to be of the same order of magnitude.}

\red{For the estimation of the absorption coefficient of the light-absorbing particle, we consider a simplified model below. The particle is made of polystyrene and contains a number of iron-oxide (Fe$_2$O$_3$/Fe$_3$O$_4$) nanoparticles at a weight fraction of $>30\%$, where the diameter of nanoparticles is estimated as $15$~nm, as shown in the scanning-electron microscopy image presented in Ref.~\cite{Paul2022a}. 
The fractions of Fe$_2$O$_3$ and Fe$_3$O$_4$ are unknown for our sample. Therefore, we consider two limiting cases: pure Fe$_2$O$_3$ and pure Fe$_3$O$_4$. Interparticle interactions between the iron-oxide nanoparticles, such as near-field coupling and multiple scattering, are neglected in the model below.}

\red{
Using the physical quantities presented in Table~\ref{tab:abs} \cite{Polyanskiy2024,Sultanova2009}, we can estimate the number density of iron-oxide nanoparticles as $\tilde{n}_{\mathrm{np}}\approx 4.5\times 10^4$ \textmu m$^{-3}$ (where the tilde distinguishes the number density from the refractive index) and the absorption cross section \cite{Jones2015} of an individual iron-oxide nanoparticle as $\sigma_{\mathrm{abs}}=16.2$~nm$^2$ for Fe$_2$O$_3$ or $\sigma_{\mathrm{abs}}=2.3$~nm$^2$ for Fe$_3$O$_4$ (see \SM~\ref{sec:SI-abs}).  These values lead to the effective absorption coefficient of the light-absorbing particle $\alpha_{\mathrm{p}}=\tilde{n}_{\mathrm{np}}\sigma_{\mathrm{abs}}=(1.0$--$7.3)\times 10^5$~m$^{-1}$. Since $\alpha_{\mathrm{p}}\gg\alpha_{\mathrm{water}},\, \alpha_{\mathrm{glass}}$, optical absorption and the resulting heat generation are expected to occur predominantly within the light-absorbing particle.}

\subsection{\red{Heat generation inside the particle}\label{sec:heatgeneration}}

\red{We estimate the heat generation rate inside the particle. Let us denote the laser power after the objective by $P_0$. For the laser power settings used in this study, $P=7.5$, $10$, $12.5$, and $15$~mW, $P_0$ was measured to be $0.35$, $0.50$, $0.61$, and $0.75$~mW, respectively, using a power meter (S405C, Thorlabs) in our optical system. 
We assume that the laser power $P_0$ is incident on the particle and that the light propagates through the particle over a distance $L(\leq d_{\rm p})$. When the laser is focused at the center of the particle, we take $L\approx d_{\rm p}$. Thus, from the definition of the absorption coefficient, the heat generation rate inside the light-absorbing particle, $Q$, is estimated as $Q=P_0(1-\exp(-\alpha_{\mathrm{p}}L))$. For instance, when $\alpha_{\rm p}=(1.0\text{--}7.3)\times10^5$~m$^{-1}$ and $P_0=0.75$~mW, $Q$ ranges from $0.095$ to $0.46$~mW. The volumetric heat generation rate is defined as $q_0=Q/V_{\rm p}$.}

\begin{table}[tb]
    \centering
    \caption{
    \red{Physical parameters used in the estimation of the absorption coefficient of the light-absorbing particle at a wavelength of $488$~nm.}}
    \begin{tabular}{l}
    \hline
        weight fraction of iron oxide \quad  
        $w=30\%$\\
        diameter of light-absorbing particle \quad  
        $d_{\mathrm{p}}=1.31$~\textmu m\\
        diameter of iron-oxide nanoparticle \quad  
        $d_{\mathrm{np}}=15$~nm\\
        mass density of polystyrene         \quad   $\rho_{\mathrm{PS}}=1.05\times10^3$ kg/m$^3$ \\
        mass density of iron oxide \quad   
        $\rho_{\mathrm{np}}=5.2\times10^3$ kg/m$^3$ \\
        refractive index of polystyrene  \quad   
        $n_{\mathrm{PS}}=1.6054$\\
        refractive index of Fe$_2$O$_3$ \quad   
        $n_{\mathrm{Fe2O3}}=3.2768$\\
        refractive index of Fe$_3$O$_4$ \quad   
        $n_{\mathrm{Fe3O4}}=2.3642$\\
        extinction coefficient of polystyrene  \quad   
        $k_{\mathrm{PS}}\approx 0$\\
        extinction coefficient of Fe$_2$O$_3$ \quad   
        $k_{\mathrm{Fe2O3}}=0.7546$\\
        extinction coefficient of Fe$_3$O$_4$ \quad   
        $k_{\mathrm{Fe3O4}}=0.0662$\\
        \hline
    \end{tabular}
    \label{tab:abs}
\end{table}

\red{\subsection{Temperature profiles\label{sec:temperature}}}
\red{
We consider a simple heat-conduction model to discuss the temperature field inside and outside the light-absorbing particle. This simplified model neglects heat conduction into the glass substrate and is intended only to provide an order-of-magnitude estimate: we assume that the particle is immersed in an infinite expanse of stationary water, where the temperature at infinity is set to $T_0$. The radius of the particle is $a_{\rm p}=d_{\rm p}/2$ and the volume of the particle is $V_{\rm p}=(4\pi/3)a_{\rm p}^3$.}

\begin{figure}[bt]
    \centering
    \includegraphics[width=\linewidth]{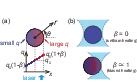}
    \caption{\red{A model of the light-absorbing particle subject to a nonuniform heat source. The glass substrate is neglected, and the particle is immersed in an infinite expanse of water. Note that the color of the particle represents the distribution of heat generation, not the temperature field.}}
    \label{fig:coordinate}
\end{figure}

\red{
We investigate the temperature field. In this section, the radial coordinate $r$ and the polar angle $\theta$ are defined such that $\theta=0$ corresponds to the heated side of the particle, as shown in Fig.~\ref{fig:coordinate}(a). Note that the axis of symmetry in the present coordinate system, $x$, is not the laser axis, as shown in Fig.~\ref{fig:coordinate}(b). We assume that the light-absorbing particle is subject to a nonuniform volumetric heat generation rate $q$. To express the nonuniform heating with minimal complexity, we prescribe $q$ as 
\begin{align}
q(r,\theta) = q_0 \left(1 + \beta \frac{r}{a_{\rm p}} \cos \theta\right), \label{eq:q}
\end{align}
where $\beta$ $(0\leq \beta \leq 1)$ is a dimensionless parameter to express the degree of bias. In this model, (i) the total heat generation rate $Q=q_0 V_{\rm p}$ is independent of $\beta$, and (ii) loose (or tight) focusing with large (or small) beam waist is represented by a small (or large) $\beta$; when $\beta=0$, we have uniform heat source $q_0$ (Fig.~\ref{fig:coordinate}(b)). 
 Solving a steady heat-conduction equation subject to the above $q$ (Eq.~\eqref{eq:q}) under the assumption of axisymmetry, we obtain the temperature field $T(r,\theta)$ (see \SM~\ref{sec:SI-temperature}), which depends on both the thermal conductivity of water $\kappa_{\rm w}$ and the thermal conductivity of the light-absorbing particle $\kappa_{\rm p}$. We use $\kappa_{\rm w}\approx 0.6$~W~m$^{-1}$~K$^{-1}$ and we assume that $\kappa_{\rm p}$ is given by the thermal conductivity of polystyrene $\approx 0.2$~W~m$^{-1}$~K$^{-1}$.} 

\red{
We evaluate three important quantities as follows:
(i) the average temperature elevation within the particle is  
 \begin{align}
\Delta T_{\rm av} = \frac{1}{2\pi}\left(1 + \frac{\kappa_{\rm w}}{5 \kappa_{\rm p}}\right)
\frac{Q}{\kappa_{\rm w} d_{\rm p}}. \label{eq:elevation}
 \end{align}
 For the present laser power setting, i.e., $Q=0.095$--$0.46$~mW, we obtain $\Delta T_{\rm av}=30.8$--$149$~K. (ii) The magnitude of temperature bias within the particle, $\Delta T_{\rm bias}(=T(a_{\rm p},0)-T(a_{\rm p},\pi))$, can be computed as
 \begin{align}
 \Delta T_{\rm bias} = \frac{3 \beta }{10 \pi }\left(1+\frac{\kappa_{\rm p}}{2 \kappa_{\rm w}}\right)^{-1}\frac{Q}{d_{\rm p} \kappa_{\rm w} }, 
 \end{align} 
 which ranges from $9.9$~K to $47.9$~K for $\beta=1$ (largely biased case). This suggests that the temperature variation within the light-absorbing particle can be substantial. (iii) Outside the light-absorbing particle, the typical magnitude of the temperature gradient, e.g., at $\theta=\pi$ (i.e., the $\mx$ heating), is 
\begin{align}
|\nabla T|_{\theta=\pi} =& \frac{1}{4\pi}\frac{Q}{a_{\rm p}^2 \kappa_{\rm w}} \left(\frac{r}{a_{\rm p}}\right)^{-2}\notag\\
&\times \left(1 - \frac{3 \beta}{5}\left(1+\frac{\kappa_{\rm p}}{2\kappa_{\rm w}}\right)^{-1}\left(\frac{r}{a_{\rm p}}\right)^{-1}\right), 
\end{align}
where the second term, which is much weaker than the first term when $r \gg a_{\rm p}$, represents the contribution of the heating bias $\beta$.
For instance, at a distance $r=5$~\textmu m, the temperature gradient is $0.47$--$2.3$~K/\textmu m, which is strong enough to induce detectable thermo-osmotic flows.}

\red{
The above estimates (i) and (iii) are expected to overestimate the temperature elevation and the magnitude of the temperature gradient in the experiment because of the absence of the glass substrate in the model, the choice of a large optical path length $L\approx d_{\rm p}$, and the assumption of the low thermal conductivity of the light-absorbing particle. For estimate (ii), the choice $\beta=1$ maximizes the temperature bias within the present model. In the experiment, when we use higher laser power, e.g., $P=20$~mW, we often encounter bubble formation, which abruptly changes the tracer motion to much faster circulating motion. Such abrupt changes are not observed in the range of $P$ investigated. The experimental observations are qualitatively consistent with the order-of-magnitude estimate of the temperature elevation given below Eq.~\eqref{eq:elevation}.}

\red{\subsection{Temperature measurement using 5CB\label{sec:5CB}}}

\red{
To estimate the overall temperature elevation of the light-absorbing particle, we follow the temperature measurement method used in Ref.~\cite{Bregulla2016}. We use a nematic liquid crystal, 4-cyano-4'-pentylbiphenyl, known as 5CB (Sigma-Aldrich), instead of water. The as-received liquid crystal is sealed in a cylindrical chamber with a glass substrate at the bottom, on which light-absorbing particles are fixed, as in the main experiment. }

\begin{figure}[bt]
    \centering
    \includegraphics[width=0.8\linewidth]{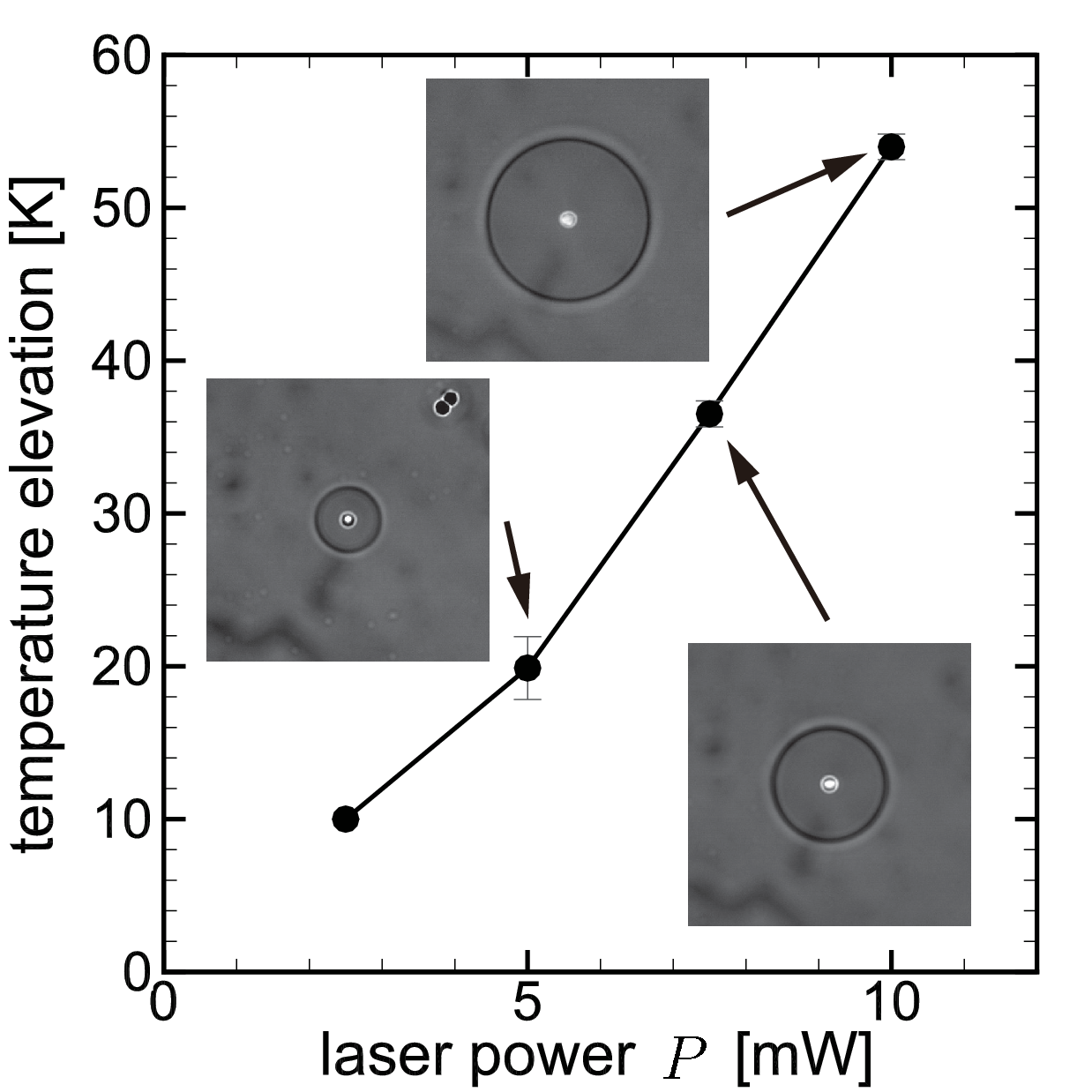}
     \caption{\red{Temperature elevation of the light-absorbing particle immersed in water, obtained by converting the temperature elevation in the liquid crystal 5CB, versus laser power $P$. The insets show typical microscope images of the phase boundaries, with the light-absorbing particle at the center of each image. For $P=2.5$~mW, the phase boundary nearly coincides with the rim of the light-absorbing particle, hindering the image analysis of $r_{\rm pb}$. Thus, we set $r_{\rm pb}=a_{\rm p}$ for $P=2.5$~mW. For $P\geq 10$~mW, the phase boundary tends to be unstable, possibly due to convection. Each data point represents the average of ten runs for different particles, and the error bars indicate the standard deviation.}}
    \label{fig:5CB}
\end{figure}

\red{
We describe the method briefly. The liquid crystal undergoes a nematic-to-isotropic phase transition at $T_{\rm pt}=308$~K. When the laser is focused on the light-absorbing particle immersed in the liquid crystal, the generated heat changes the phase of the liquid crystal only near the particle. The phase boundary can be easily observed by bright-field microscopy, as shown in the insets of Fig.~\ref{fig:5CB}. Assuming that the particle is immersed in an infinite expanse of liquid crystal, we obtain the temperature field outside the particle as $T_{\rm 5CB}(r)=T_0+\Delta T_{\rm 5CB}/(r/a_{\rm p})$, where $\Delta T_{\rm 5CB}$ is the temperature elevation at the particle surface. Therefore, once the radius $r_{\rm pb}$ of the phase boundary is determined from the microscope image, we can compute the temperature elevation at the particle surface as $\Delta T_{\rm 5CB}=(T_{\rm pt}-T_0)(r_{\rm pb}/a_{\rm p})$. } 

\red{
The temperature elevation measured in 5CB can be converted to that in water by multiplying it by $\kappa_{\rm 5CB}/\kappa_{\rm w}$, where $\kappa_{\rm 5CB}=0.15$~W~m$^{-1}$~K$^{-1}$ is the thermal conductivity of the liquid crystal. This follows from the inverse proportionality between the temperature elevation outside the particle and the thermal conductivity of the surrounding fluid for $\beta=0$ (see Eq.~\eqref{eq:SI-temp2} with $\beta=0$ for water). The converted temperature elevation is plotted in Fig.~\ref{fig:5CB} using $T_0=293$~K. The temperature elevation shows an increasing trend with $P$, which slightly deviates from a linear function. Notably, the order of magnitude is consistent with the prediction in Appendix~\ref{sec:temperature}. Therefore, we consider that this measurement captures the qualitative trend of the temperature elevation with increasing laser power. For further quantification, the model can be improved to include, e.g., the effect of the substrate and the temperature dependence of physical properties. These effects may explain the slight nonlinearity observed in Fig.~\ref{fig:5CB} and the discrepancy between the prediction and the experimental results. }

\begin{figure}[bt]
    \centering
    \includegraphics[width=1\linewidth]{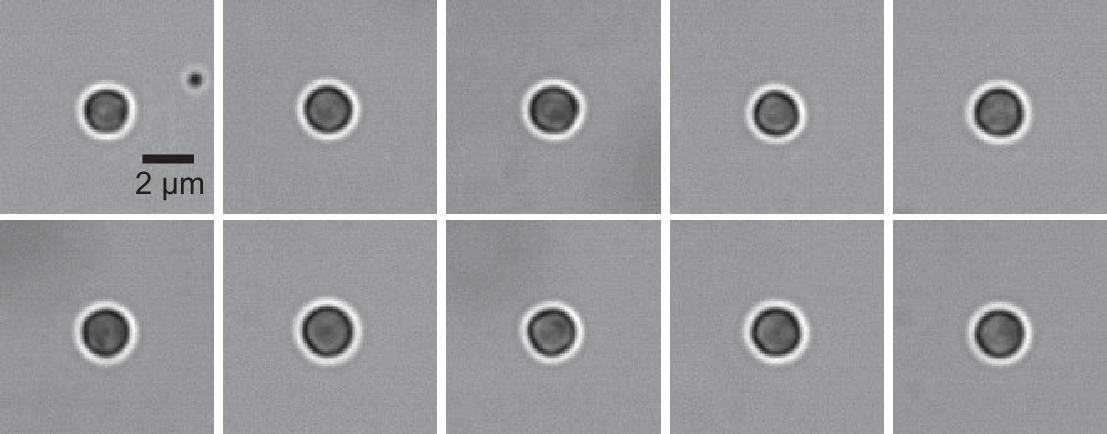}
    \caption{\red{Typical bright-field images of the light-absorbing particles. The shapes and internal textures vary slightly among individual particles.}}
    \label{fig:particle-image}
\end{figure}

\red{
\section{Bright-field image analysis of the light-absorbing particles\label{sec:particle}}
}
\red{
To examine the homogeneity of the light-absorbing particles, we show typical bright-field images of the particles in Fig.~\ref{fig:particle-image}, with the focal plane set at the particle equator. The cross-sectional shapes of the individual particles are seen to deviate slightly from a perfect circle. We also note that the internal textures vary among individual particles. 
}

\red{
To further characterize the particle-to-particle variation, we analyze the equivalent circular diameters $d_{\rm p}^{\ast}$ of the ten particles shown in Fig.~\ref{fig:particle-image}. Excluding the white rim surrounding each particle, we obtain $d_{\rm p}^{\ast}=1.38\pm0.06$~\textmu m, which is close to the catalog value of $d_{\rm p}=1.31\pm0.06$~\textmu m. The aspect ratio is $1.03\pm0.01$, indicating a slight deviation from a perfect circular shape. These observations indicate that the light-absorbing particles are not perfectly homogeneous, and the run-to-run scatter observed, for example, in Fig.~\ref{fig:11} may partly originate from such particle-to-particle variations.
}

\begin{figure*}[bt]
    \centering
    \includegraphics[width=1\linewidth]{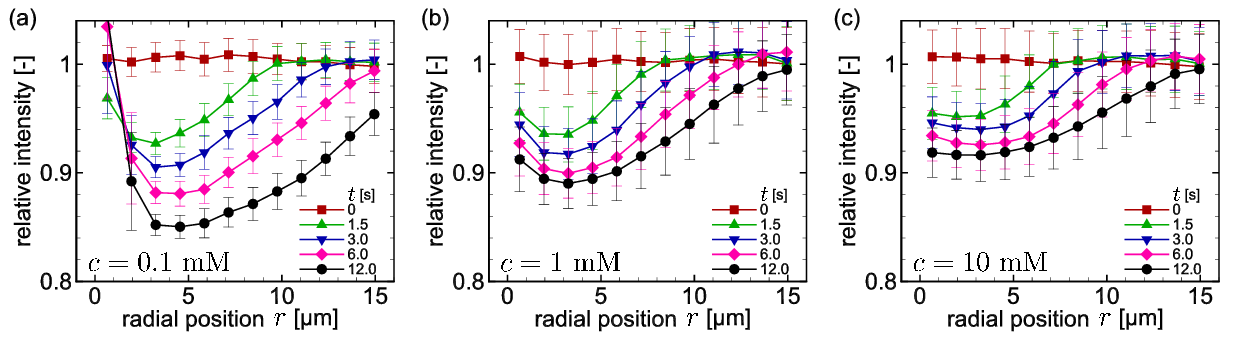}
    \caption{\red{Relative intensity distributions, normalized by the spatially averaged intensity at the initial time, are plotted as a function of the radial position $r$ for (a) $c=0.1$~mM, (b) $c=1$~mM, and (c) $c=10$~mM and for $t=0$, $1.5$, $3$, $6$, and $12$~s.}}
    \label{fig:radial-intensity}
\end{figure*}

\red{
\section{Electrolyte effect on tracer thermophoresis\label{sec:thermophoresis}}
}

\red{The surfaces of the tracers in this study are carboxylate-modified and negatively charged in tris-HCl solution \cite{Tsuji2018b}. To investigate the effect of electrolyte concentration on tracer thermophoresis, we carry out auxiliary experiments as follows. We fabricate the same fluidic chamber as in the main experiments, but we do not introduce light-absorbing particles. Rather, we use an infrared laser with a wavelength of $1480$~nm to heat the water directly \cite{Duhr2006,Tsuji2021,Tsuji2023}. In this way, we aim to produce a temperature distribution that is more uniform in the $z$ direction. The laser is loosely focused at $z=5$~\textmu m, the midplane of the device, through an objective with high transmittance at $1480$~nm (NA=0.85, LCPLN100XIR, Olympus). Under this setting, we observe the radial thermophoretic motion of the tracers for $c=0.1$, $1$, and $10$~mM. Note that, for this heating method, the lower-NA objective does not allow sufficiently tight optical trapping, and thus we cannot use ot-PTV.}

\red{The time evolution of the intensity distribution of the tracers is presented in Fig.~\ref{fig:radial-intensity} as a function of $r$, the distance from the laser axis. A low-intensity region develops around the laser axis and expands rapidly over time, extending beyond $r=5 $~\textmu m in all cases, indicating thermophobic (from hot to cold) behavior. This direction is consistent with the literature, e.g., Ref.~\cite{Duhr2006}. }

\red{We observe two distinct trends depending on $c$. First, at $c=0.1$~mM, some tracers are trapped near the center, as indicated by the increase in intensity at $r=0$~\textmu m. The trapped tracers appear to remain near the bottom substrate. We consider that this trapping occurs due to the thermo-osmotic flow near the bottom substrate \cite{Tsuji2024}. This phenomenon is also the subject of ongoing research in our group. Second, the tracer depletion is more significant for lower $c$, suggesting that tracer thermophoresis is stronger at lower electrolyte concentrations. This trend is consistent with previous thermophoresis experiments, such as those reported in Ref.~\cite{Duhr2006}. }

\vspace{1em}
\begin{acknowledgments}
This work was supported by the Japan Society for the Promotion of Science KAKENHI Grants 
No.~JP24K00803, 
No.~JP24K00822, 
No.~JP24H00293, 
No.~JP25K01156, 
No.~JP25K22071, 
No.~JP25KJ1967, 
and also by the Japan Science and Technology Agency PRESTO Grant 
No.~JPMJPR22O7, 
No.~JPMJPR23O8, 
and FOREST Grant No.~JPMJFR2445. 
The authors thank Mr. Mitsuki Ito and Takumi Kitano for the assistance in the experiment.
\end{acknowledgments}

\section*{Data availability statement}
The plotting data presented in this article will be openly available \cite{TST2026-DB} when published. Detailed data such as source codes are available from the corresponding author upon reasonable request.

%


\clearpage

\appendix
\onecolumngrid

\setcounter{page}{1}
\renewcommand{\appendixname}{\SM}

\begin{center}
\SM~on \\[1em]
{\large
{\bf \titleB}}
\\[1em]
{\small
Tetsuro Tsuji$^{\dagger,\sharp,\ast}$, Shota Suzuki$^\dagger$, Satoshi Taguchi$^\dagger$, Haruya Ishida$^\ddagger$, and Hideaki Teshima$^{\ddagger,\S}$\\[0.5em]
$^\dagger$Department of Informatics, Kyoto University, Kyoto 606-8501, Japan\\[0.25em]
$^\ddagger$Department of Aeronautics and Astronautics, Kyushu University, Fukuoka 819-0395, Japan\\[0.25em]
$^\S$International Institute for Carbon-Neutral Energy Research (WPI-I2CNER), Kyushu University, Fukuoka 819-0395, Japan\\[0.25em]
$^\sharp$Present address: Department of Aeronautics and Astronautics, Kyoto University, Kyoto 615-8540, Japan\\[0.25em]
$^\ast$tsuji.tetsuro.7x@kyoto-u.ac.jp (corresponding author); 
}
\end{center}


\setcounter{figure}{0}
\renewcommand\thefigure{S\arabic{figure}}   
\setcounter{table}{0}
\renewcommand\thetable{S\arabic{table}} 
\renewcommand\theequation{\thesection.\arabic{equation}} 
\renewcommand{\thesection}{S\arabic{section}}
\renewcommand{\thesubsection}{\Alph{subsection}}
\renewcommand{\thesubsubsection}{\Roman{subsubsection}}

\section{Movie}\label{sec:SM-movie}
\begin{itemize} 
\item \verb/direction-control.mp4/ is the movie file corresponding to Fig.~\ref{fig:exp}(b,c,d,e).
\item \verb/ot-PTV.mp4/ is an example of the present ot-PTV experiment of the $\mx$ heating case. The distance between the heating position, i.e., the left-edge of the light-absorbing particle, and the measurement position is set to $5$~\textmu m (i.e., $r=5$~\textmu m and $\theta=0$). 
\end{itemize}

\section{Thermo-osmotic coefficient based on Derjaguin's model}\label{sec:SI-theory}

From \red{a} macroscopic point of view, thermo-osmotic flow can be \red{incorporated} into the analysis of fluid motion by imposing a thermal-slip boundary condition on a boundary. In the literature, Derjaguin's model \cite{Derjaguin1987} has been widely applied, that is, 
\begin{align}
\bm{v}\cdot \bm{t} = \chi \frac{\nabla T}{T}\cdot \bm{t} \quad (\text{on boundary}), \quad \chi = -\frac{1}{\eta} \int_{0}^\infty z h(z) \rd z, \label{eq:SI-derjaguin}
\end{align}
where $\bm{v}$ is the flow velocity, $\bm{t}$ is the unit tangent \red{to} the boundary, $\bm{v}\cdot \bm{t}$ is the thermal-slip velocity, $\chi\red{(=-K T)}$ is the thermo-osmotic coefficient, $T$ is the temperature of the fluid, $z$ is the distance from the boundary, $\eta$ is the fluid viscosity, and $h$ is the excess enthalpy density (i.e., the deviation of enthalpy \red{density} from \red{its bulk} value). Note that \red{the variable} $z$ here indicates a boundary-layer coordinate, and thus $z\to\infty$ represents the bulk. The quantity $h$ is the ``excess" value, and thus is denoted by $\delta h$ in some references, but we omit ``$\delta$" here for concise notation. 

The expression of $h$ is crucial in this model. In the following, we summarize the calculation of $h$, referring to Refs.~\cite{Fraenzl2022,Herrero2022}. The excess enthalpy arises due to various physical \red{effects}, such as electrostatic effect\red{s}, intermolecular potential\red{s} between fluid molecules and the boundary (or solid atoms), the dipole moment of the solvent induced by the surface charge of the boundary, and so on. Refs.~\cite{Fraenzl2022,Herrero2022} \red{suggest} that electrostatic effect\red{s} and intermolecular \red{interactions} are \red{the} dominant \red{contributions}. In the following, we \red{re}view these two \red{contributions}. 

\subsection{Electrostatic effect}

\subsubsection{Formulation}

Let us consider the steady state of an electrolyte solution \red{in} an electric potential \red{field} $\psi(\bx)$. Let $n_+(\bx)$ and $n_-(\bx)$ denote the concentrations of monovalent cations and anions, respectively. Furthermore, the dielectric permittivity $\varepsilon$ of a solvent (i.e., water) is considered constant for simplicity. Note that the temperature-dependent dielectric permittivity was treated in Refs.~\cite{Fayolle2008,Fraenzl2022}, while its contribution \red{to thermo-osmotic flow was found to be} limited. 

In thermodynamic equilibrium, uniform electrochemical potential\red{s} result in the distributions $n_\pm$ obeying the Boltzmann distribution: 
\begin{align}
n_{\pm} = n_0 \exp(\mp \beta e \psi), \quad \beta = \frac{1}{\kB T}, 
\end{align}
where $n_0$ is a reference density at zero potential (e.g., in the bulk), $\beta$ is \red{the} inverse temperature, and $e$ is the elementary charge. These distributions are coupled with the Poisson equation in a self-consistent manner, that is, 
\begin{align}
\Delta \psi = -\frac{\rhoe}{\varepsilon}, \quad \rhoe = e(n_+-n_-)=-2n_0e\sinh(\beta e \psi), \label{eq:SI-pb}
\end{align}
which is known as the Poisson--Boltzmann equation. 

In the following, we consider a simple case near a plane boundary at $z=0$, assuming $\psi=\psi(z)$. Furthermore, temperature $T$ (and thus $\beta$) is not dependent on $z$, since the length scale of temperature variation is much larger than the Debye length $\db$ introduced below, which determines the length scale of $\psi$. The potential is assumed to decay at infinity, that is, $\psi\to0$ as $z\to\infty$. The other boundary condition at $z=0$ can be either (i) a fixed potential $\psi=\psi_0$ at $z=0$, (ii) a fixed surface charge density $\sigma$, or \red{a combination} of (i) and (ii). It should be noted that Gauss's law reads 
\begin{align}
\sigma = - \varepsilon \left.\diff{\psi}{z}\right|_{z=0}, \label{eq:SI-gauss}
\end{align}
on the surface\red{. Thus, for a given $\sigma$,} condition (ii) \red{corresponds to a} Neumann-type boundary condition for $\psi$. To summarize, the boundary conditions for $\psi$ \red{are} given \red{by} 
\begin{subequations}\label{eq:SI-bc}
\begin{align}
&\text{case (i):\quad} \psi = \psi_0 \quad(z=0), \quad \psi\to0 \quad (z\to\infty). \label{eq:SI-bc1}\\
&\text{case (ii):\quad}\sigma = -\varepsilon \left.\diff{\psi}{z}\right|_{z=0} \quad(z=0), \quad \psi\to0 \quad (z\to\infty). \label{eq:SI-bc2}
\end{align}
\end{subequations}

Once the potential $\psi$ is obtained, the excess enthalpy \red{density} $h$ can be expressed in terms of $\psi$ as follows. The excess internal energy density of an electrolyte solution is $\rhoe\psi$, and thus the excess enthalpy density is 
\begin{align}
h=\rhoe\psi + p, \label{eq:SI-h}
\end{align} 
where $p$ is the deviation of the pressure from \red{its bulk value}. \red{Under the assumption of} local thermal equilibrium, we consider $h$ as \red{a} function of $z$. The \red{balance} of forces acting \red{on} the fluid in the $z$ direction leads to 
\begin{align}
- \diff{p}{z} - \rhoe \diff{\psi}{z} = 0. 
\end{align}
Using the Poisson equation $\diff{^2 \psi}{z^2}=-\frac{\rhoe}{\varepsilon}$, we calculate $p$ as
\begin{align}
\diff{p}{z} = \varepsilon \diff{\psi}{z} \diff{^2\psi}{z^2} = \frac{\varepsilon}{2}\diff{}{z}\left(\rnd{\psi}{z}\right)^2 \quad \Longrightarrow \quad
p=\frac{\varepsilon}{2}\left(\diff{\psi}{z}\right)^2, 
\end{align}
where we set the integration constant to zero by considering $p\to0$ and $\diff{\psi}{z}\to0$ as $z\to \infty$. Therefore, the \red{excess} enthalpy \red{density} \eqref{eq:SI-h} is expressed as 
\begin{align}
h = -\varepsilon \psi \diff{^2\psi}{z^2} + \frac{\varepsilon}{2} \left(\diff{\psi}{z}\right)^2. \label{eq:SI-h2}
\end{align}

Before going into detail, let us introduce dimensionless quantities as 
\begin{equation}
\begin{split}
&\tz = \frac{z}{\db} \quad \left(\db =\sqrt{\frac{\varepsilon}{2 e^2 n_0 \beta}}\right), \quad  \tpsi(\tz) = \beta e \psi(z), \quad 
\tpsi_0=\beta e \psi_0,\\
&\th = \frac{h}{\varepsilon / (\beta^2 e^2 \db^2)}
= \frac{h}{2n_0/\beta}, \quad \tchi = \frac{\chi}{\varepsilon/(\eta e^2 \beta^2)}, \quad 
\tsigma = \frac{\sigma}{\varepsilon/(\db \beta e)}=\frac{\sigma}{\sqrt{2n_0\varepsilon/\beta}},
\end{split}
\end{equation}
where $\db$ is the Debye length. Then, Eqs.~\eqref{eq:SI-pb}, \eqref{eq:SI-bc}, the excess enthalpy \eqref{eq:SI-h2}, \red{and} the thermo-osmotic coefficient \eqref{eq:SI-derjaguin} are recast as
\begin{subequations}\label{eq:SI-pb-nd}
\begin{align}
&\text{Poisson--Boltzmann Equation:\quad}\diff{^2 \tpsi}{\tz^2} = \sinh(\tpsi), \\
&\text{case (i):\quad }\tpsi = \tpsi_0 \quad(\tz=0), \quad \tpsi\to0 \quad (\tz\to\infty), \label{eq:SI-constant-potential}\\
&\text{case (ii):\quad }\tsigma = -\rnd{\tpsi}{\tz} \quad(\tz=0), \quad \tpsi\to0 \quad (\tz\to\infty), \label{eq:SI-constant-charge}\\
&\th = -\tpsi \rnd{^2\tpsi}{\tz^2} + \frac{1}{2}\left(\rnd{\tpsi}{\tz}\right)^2, \quad 
\tchi=-\int_0^\infty \tz \th (\tz) \rd \tz. 
\end{align}
\end{subequations}

\subsubsection{Analysis of the Poisson--Boltzmann equation}

Firstly, let us consider case (i), Eq.~\eqref{eq:SI-constant-potential}. The Poisson--Boltzmann equation \eqref{eq:SI-pb} has \red{the} so-called G\red{ou}y--Chapman (GC) solution: 
\begin{align}
\tpsi(\tz) = \tpsi_\GC(\tz) \equiv 4\,\arctanh(\gamma\exp(-\tz)) \left(=2 \ln (\frac{1+\gamma \exp(-\tz)}{1-\gamma \exp(-\tz)})\right), \quad \gamma = \tanh(\frac{\tpsi_0}{4}), \label{eq:SI-gc}
\end{align}
In particular, when $|\tpsi|\ll1$ (say, $|\tpsi_0|\ll1$\red{,} or \red{equivalently} $\red{|}e\psi_0\red{|} \ll \kB T$), Eq.~\eqref{eq:SI-pb-nd} is reduced to $\rnd{^2\tpsi}{\tz^2}=\tpsi$, which is called the Debye--H\"uckel (DH) approximation. This case has a simpler analytical form: 
\begin{align}
\tpsi(\tz) = \tpsi_\DH(\tz) \equiv \tpsi_0\exp(-\tz),\quad 
\th=-\frac{\tpsi_0^2}{2}\exp(-2\tz). \label{eq:SI-dh}
\end{align}
For these two cases, the \red{thermo-osmotic coefficient} is calculated explicitly as 
    \begin{align}
        \tchi = 
        \begin{cases}
        \dfrac{\tpsi_0^2}{2}+3\tpsi_0-12\ln(\dfrac{\exp(\tpsi_0/2)+1}{2}) \quad (\text{for }\tpsi=\tpsi_\GC), \\[1em]
        \dfrac{\tpsi_0^2}{8} \quad (\text{for }\tpsi=\tpsi_\DH), \\
        \end{cases}
        \label{eq:SI-chi}
    \end{align}
where the first line \red{reduces} to the second line when $|\tpsi_{\red{0}}|\ll1$. In dimensional form, we have  
    \begin{align}
        \chi = 
        \begin{cases}
        \dfrac{\varepsilon}{\eta}\left[\dfrac{\psi_0^2}{2}+3\dfrac{\psi_0}{\beta e}-\dfrac{12}{\beta^2 e^2}\ln(\dfrac{\exp(\beta e \psi_0/2)+1}{2})\right] \quad (\text{for }e\psi_0 \sim \kB T), \\[1em]
        \dfrac{\varepsilon \psi_0^2}{8 \eta} \quad (\text{for } \red{|}e\psi_0\red{|}\ll\kB T). \\
        \end{cases}
    \end{align}

\begin{figure}
    \centering
    \includegraphics[width=0.6\linewidth]{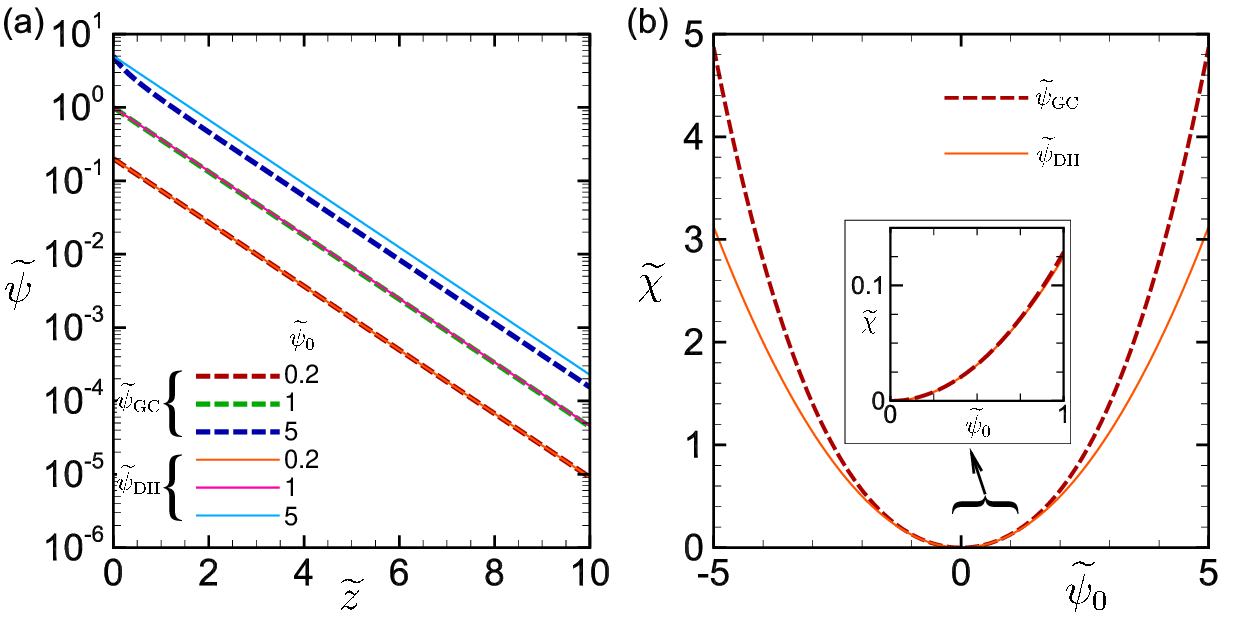}
    \caption{(a) Solution of the Poisson--Boltzmann equation $\tpsi=\tpsi_\GC$ (dashed) and $\tpsi=\tpsi_\DH$ (solid) for the surface potential $\tpsi_0=0.2$, $1$, and $5$ (i.e., case (i)). (b) Thermo-osmotic coefficient $\tchi$ as a function of the surface potential $\tpsi_0$ for $\tpsi=\tpsi_\GC$ (dashed) and $\tpsi=\tpsi_\DH$ (solid). }
    \label{fig:SI-theory}
\end{figure}

Figure~\ref{fig:SI-theory}(a) shows Eqs.~\eqref{eq:SI-gc} and \eqref{eq:SI-dh} for $\tpsi_0=0.2$ (weak potential), $1$ (moderate potential), and $5$ (strong potential), where $\tpsi$ decays exponentially as $\tz$ \red{increases}. For $\red{|}\tpsi_{\red{0}}\red{|}\leq 1$, the results of the GC solution and the DH approximation are in quantitative agreement. Figure~\ref{fig:SI-theory}(b) shows the (dimensionless) thermo-osmotic coefficient $\tchi$ \red{in} Eq.~\eqref{eq:SI-chi} as \red{a} function of the surface potential $\tpsi_0$. It is seen that $\tchi$ is an increasing function of $|\tpsi_0|$, and the GC solution and the DH approximation are in good agreement for $\red{|}\tpsi_{\red{0}}\red{|}\leq 1$. \red{Under} typical experimental condition\red{s} at room temperature, e.g., $\varepsilon= 80.4\varepsilon_0$ ($\varepsilon_0=8.85\times10^{-12}$~F/m is the \red{vacuum} permittivity), $\eta=1.0\times10^{-3}$~Pa~s, $\kB=1.38\times10^{-23}$~J/K, $e=1.60\times10^{-19}$~C, and $T=300$~K, the reference potential is $\psi_{\mathrm{ref}}=(\beta e)^{-1}=25.9$~mV and the reference thermo-osmotic coefficient is $\chi_{\mathrm{ref}}=\varepsilon/(\eta e^2 \beta^2)=4.76\times 10^{-10}$~m$^2$/s. Figure~\ref{fig:SI-theory}(a) is a good starting point to see how $\tpsi$ behaves \red{for different values of} $\tpsi_0$. However, $\tpsi_0$ \red{cannot be controlled} in our \red{experimental} setup.  

Next, let us consider case (ii), Eq.~\eqref{eq:SI-constant-charge}. The G\red{ou}y--Chapman solution \eqref{eq:SI-gc}, i.e., $\tanh (\tpsi/4)=\gamma \exp(-\tz)$, leads to 
\begin{align}
\left.\rnd{\tpsi}{\tz}\right|_{\tz=0} = -2\sinh(\frac{\tpsi_0}{2}).
\end{align}
Substituting this into Eq.~\eqref{eq:SI-constant-charge}, we obtain the (dimensionless) Grahame equation
\begin{align}
\tsigma = 2 \sinh(\frac{\tpsi_0}{2}) \quad \longrightarrow \quad \tpsi_0=2\arcsinh(\frac{\tsigma}{2}), 
\end{align}
the dimensional form of which is 
\begin{align}
\sigma = \frac{2\varepsilon}{e \beta \db} \sinh(\frac{e\beta \psi_0}{2})=\sqrt{8\varepsilon n_0 \kB T}\sinh(\frac{e\psi_0}{2\kB T}).  
\end{align}
Therefore, when we consider case (ii), we specify $\tsigma$ to obtain $\tpsi_0(=2\arcsinh(\frac{\tsigma}{2}))$. 

\red{Finally, we need to relate $n_0$ to the electrolyte concentration $c$ in the experiment. We note that the ion density $n_0$ is related to the ionic strength $I_{\rm ion}$ as $n_0=I_{\rm ion}N_A$, where $N_A$ is the Avogadro number. For the present experimental conditions with a tris-HCl solution at $\rm pH=8.0$, $I_{\rm ion}$ is computed as $I_{\rm ion}=0.53 \times c$ using the value of $\rm pKa = 8.06$ at room temperature. This means that approximately $53$~\% of the tris molecules are protonated.} 

For the \red{constant-charge condition} of Eq.~\eqref{eq:SI-constant-charge}, the value of $\sigma$ is difficult to predict. Nonetheless, referring \red{to} Ref.~\cite{Behrens2001}, where the effective surface charge \red{was} theoretically \red{estimated}, we set $\sigma=-0.32$~mC/m$^2$ in the following as a reference case. Then, $\tsigma=\sigma/\sqrt{2n_0\varepsilon \kB T}$ and the corresponding surface potential are computed as 
\begin{equation}
    \begin{split}
        & \tsigma = \red{-0.073} \quad \Longrightarrow \quad \psi_0 = \red{-1.9}~\text{mV}, \quad (c=10~\text{mM}),\\
        & \tsigma = \red{-0.23} \quad \Longrightarrow \quad \psi_0 = \red{-6.0}~\text{mV}, \quad (c=1~\text{mM}),\\
        & \tsigma = \red{-0.74} \quad \Longrightarrow \quad \psi_0 = -\red{18.4}~\text{mV}, \quad (c=0.1~\text{mM}),\\
    \end{split}
\end{equation}
where the values of $c$ in the main text are used. These values of $\psi_0$ are presented in Table~\ref{tab:AFM}.

\subsection{Effect of intermolecular potential}
For the inter-molecular potential, Ref.~\cite{Herrero2022} computed $h$ directly from molecular dynamic\red{s} simulation\red{s}. Therefore, it is not possible to apply their results to the present case, where the molecular details are different. The van der Waals potential between fluid molecules and the surface is considered in Ref.~\cite{Fraenzl2022}, based on a phenomenological fluid-dynamic model. The model is rather simple but contains phenomenological parameters such as the size of solvent molecules $d_0$ and the Hamaker constant $A_H$. These parameters are difficult to determine experimentally, and thus remain ambiguous. 

\red{When the surface is coated} with Pluronic F-127, which is a nonionic copolymer, the electrostatic effect is \red{expected to be} reduced and \red{could be} negligible. \red{On the other hand}, the hydrophilic polyethylene-glycol part of Pluronic F-127 is expected to cover the glass surface, modulating the excess enthalpy profile. According to Ref.~\cite{Bregulla2016}, the thermo-osmotic coefficient for \red{a} Pluronic\red{-}coat\red{ed surface} is estimated \red{to be} $\chi=14~\times10^{-10}$~m$^2$/s from a simple rod-like model of polymers. We expect a similar magnitude of the thermo-osmotic coefficient here. The electrostatic contribution to $\chi(=\tchi \chi_{\mathrm{ref}})$ \red{satisfies} $|\tchi|<0.1$ $(|\chi|<5\times10^{-11}$~m$^2$/s) for $\chi_{\mathrm{ref}}=4.76\times10^{-10}$~m$^2$/s, as presented in Fig.~\ref{fig:SI-theory}(b), and thus the contribution \red{from} intermolecular \red{interactions} ($\chi=14~\times10^{-10}$~m$^2$/s) is \red{much larger}, as \red{suggested by} the case of $\mx$ heating in Fig.~\ref{fig:add}. 

\section{\red{Estimation of the absorption coefficient of a light-absorbing particle}}\label{sec:SI-abs}

\red{First, we compute the number density of the iron-oxide nanoparticles within the light-absorbing particle. Using the mass density values in Table~\ref{tab:abs}, the volume fraction of iron oxide is 
\begin{align}
w_{\mathrm{v}}= \frac{(w/\rho_{\mathrm{np}})}{(w/\rho_{\mathrm{np}})+(1-w)/\rho_{\mathrm{PS}}}\approx 8\%, 
\end{align}
which leads to the total number of nanoparticles $N_{\mathrm{np}}=V_{\mathrm{p}}w_{\mathrm{v}} / V_{\mathrm{np}}$, where $V_{\mathrm{p}}$ and $V_{\mathrm{np}}$ are the volumes of the light-absorbing particle and the nanoparticle, respectively. This further gives the number density of nanoparticles $\tilde{n}_{\mathrm{np}}=N_{\mathrm{np}}/V_{\mathrm{p}}=w_{\mathrm{v}} / V_{\mathrm{np}}$, which is presented in the main text.}

\red{Next, we compute the absorption cross section. 
We consider the iron-oxide nanoparticles with a diameter of $d_{\mathrm{np}}=15$~nm irradiated by light with a wavelength of $\lambda = 488$~nm. Then, noting that $d_{\mathrm{np}}/\lambda\ll1$, we can use the Rayleigh theory as the leading-order approximation. That is, we can assume that the polarizability of a single nanoparticle in a background medium (i.e., polystyrene), $\tilde{\alpha}_{\mathrm{np}}$, is approximated by the Clausius--Mossotti relation $\tilde{\alpha}_{\mathrm{CM}}$ \cite{Jones2015} (the tilde distinguishes the polarizability from absorption coefficients)
\begin{align}
\tilde{\alpha}_{\mathrm{np}}\approx \tilde{\alpha}_{\mathrm{CM}} = 4 \pi \varepsilon_0 \varepsilon_{\mathrm{PS}} \left(\frac{d_{\mathrm{np}}}{2}\right)^3\frac{\varepsilon_{\mathrm{np}}-\varepsilon_{\mathrm{PS}}}{\varepsilon_{\mathrm{np}}+2\varepsilon_{\mathrm{PS}}},
\end{align}
where $\varepsilon_{\mathrm{PS}}=n_{\mathrm{PS}}^2$ is the relative permittivity of polystyrene, $\varepsilon_{\mathrm{np}}=(n_{\mathrm{np}}+ \mathrm{i}\,k_{\mathrm{np}})^2$ is the relative permittivity of an iron-oxide nanoparticle, and $\mathrm{i}$ is the imaginary unit. For the values of $n_{\mathrm{np}}$ and $k_{\mathrm{np}}$, we use the values of $n_{\mathrm{x}}$ and $k_{\mathrm{x}}$ with either $\mathrm{x}=$Fe$_2$O$_3$ or Fe$_3$O$_4$ in Table~\ref{tab:abs}. Then, we can compute the extinction cross section as \cite{Jones2015} 
\begin{align}
\sigma_{\mathrm{ext}}=\mathrm{Im}\left[\frac{k_{\mathrm{PS}}}{\varepsilon_0 \varepsilon_{\mathrm{PS}}}\tilde{\alpha}_{\mathrm{CM}}\right]\quad \left(k_{\mathrm{PS}} = \frac{2 \pi n_{\mathrm{PS}}}{\lambda}\right), 
\end{align}
where $\mathrm{Im[\cdot]}$ indicates the imaginary part, and $k_{\mathrm{PS}}$ is the wavenumber in polystyrene. 
Usually, the extinction cross section is the sum of the scattering cross section $\sigma_{\mathrm{sca}}$ and absorption cross section $\sigma_{\mathrm{abs}}$, but we can neglect the scattering cross section $(\sigma_{\mathrm{sca}}\propto d_{\mathrm{np}}^6)$ compared with the absorption cross section $(\sigma_{\mathrm{abs}}\propto d_{\mathrm{np}}^3)$ in the Rayleigh theory. Thus, we obtain $\sigma_{\mathrm{abs}}\approx \sigma_{\mathrm{ext}}$, which can be computed using the values of Table~\ref{tab:abs}. Finally, the effective absorption coefficient of the light-absorbing particle is estimated as $\alpha_{\rm p}=\tilde{n}_{\mathrm{np}}\sigma_{\mathrm{abs}}$.}

\section{\red{Details of the temperature field in Appendix~\ref{sec:temperature}}\label{sec:SI-temperature}}

\red{We consider the steady-state heat conduction equation subject to a heat source $q$ (Eq.~\eqref{eq:q}) inside the light-absorbing particle of radius $a_{\rm p}$ in an infinite expanse of water. The spherical coordinates $(r,\theta)$ are defined as shown in Fig.~\ref{fig:coordinate}. The boundary conditions at $r=a_{\rm p}$ require continuity of temperature $T(r,\theta)$ and heat flux; the condition at infinity is $T\to T_0$. With this simplified setting, we can obtain an explicit temperature field as shown below.}

\red{ 
The temperature field $T(r,\theta)$ inside the light-absorbing particle is obtained as
\begin{align} 
\frac{T(r,\theta) - T_ 0}{T_0} = 
\frac{\Delta T_{\rm ref}}{T_0}\left\{
1+\frac{\kappa_{\rm w}}{2\kappa_{\rm p}}-\frac{\kappa_{\rm w}}{2\kappa_{\rm p}}\left(\frac{r}{a_{\rm p}}\right)^2 + \frac{3}{10} \frac{\kappa_{\rm w}}{\kappa_{\rm p}}\left[\frac{3+2 (\kappa_{\rm w}/\kappa_{\rm p})}{1+2 (\kappa_{\rm w}/\kappa_{\rm p})}
\frac{r}{a_{\rm p}}-\left(\frac{r}{a_{\rm p}}\right)^{3}\right]\beta \cos\theta
\right\} \quad (r\leq a_{\rm p}), \label{eq:SI-temp1}
\end{align} 
where $\Delta T_{\rm ref} = (q_0 a_{\rm p}^2)/(3 \kappa_{\rm w} )$ is the reference temperature elevation for infinite thermal conductivity of the particle (i.e., $\kappa_{\rm p}\gg \kappa_{\rm w}$). The temperature outside the light-absorbing particle is obtained as
\begin{align} 
\frac{T(r,\theta) - T_ 0}{T_0} = 
\frac{\Delta T_{\rm ref}}{T_0}\left\{
\left(\frac{r}{a_{\rm p}}\right)^{-1}+ \frac{3}{10} \left(1+\frac{\kappa_{\rm p}}{2 \kappa_{\rm w}}\right)^{-1}\left(\frac{r}{a_{\rm p}}\right)^{-2}\beta \cos\theta
\right\} \quad (r > a_{\rm p}). \label{eq:SI-temp2} 
\end{align}
}

\begin{figure}
    \centering
    \includegraphics[width=0.7\linewidth]{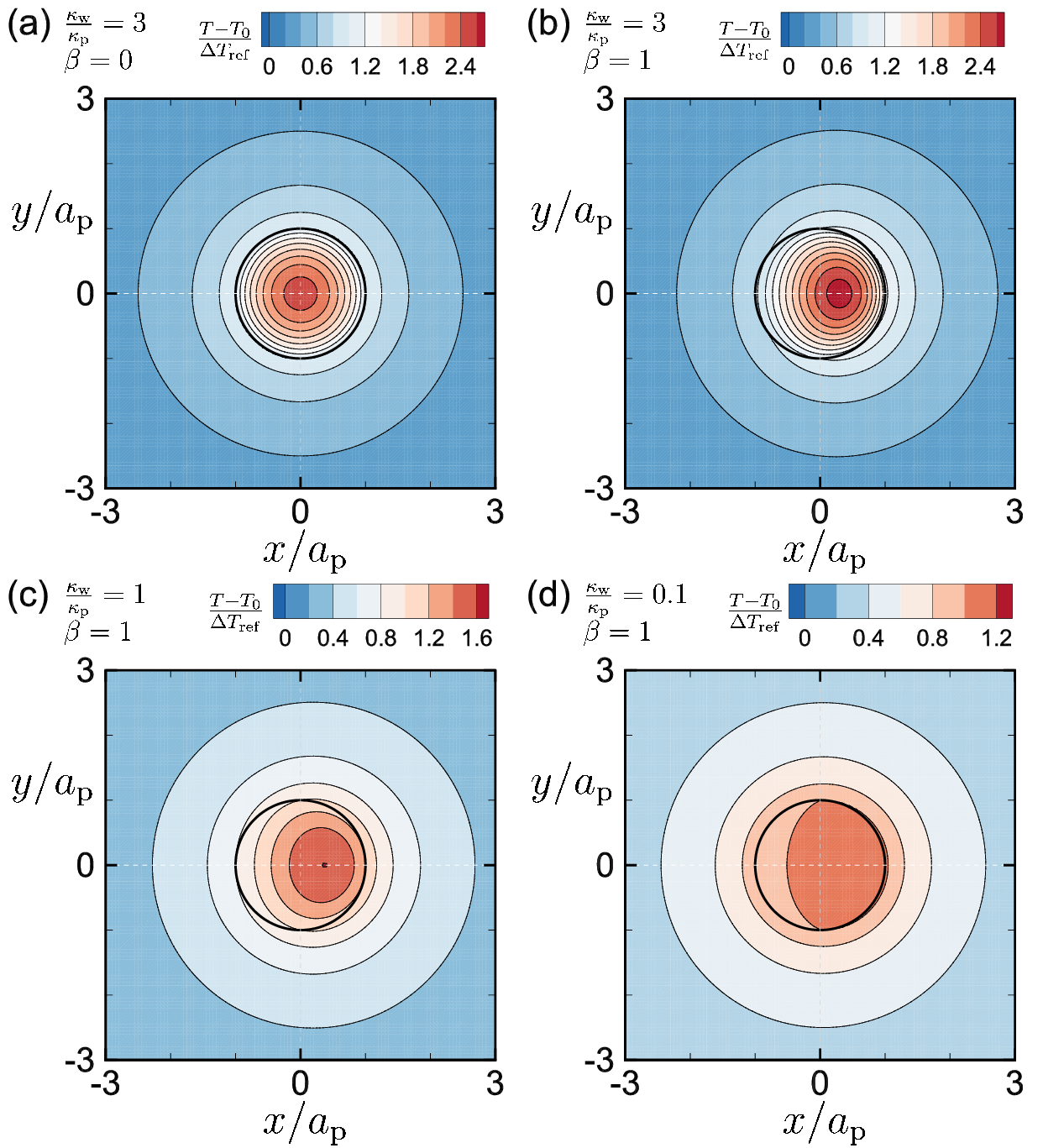}
    \caption{\red{Contour plots of the temperature field $(T-T_0)/\Delta T_{\rm ref}$, i.e., the expressions enclosed in braces in Eqs.~\eqref{eq:SI-temp1} and \eqref{eq:SI-temp2} with $x=r\cos\theta$ and $y=r\sin\theta$. (a) $\kappa_{\rm w}/\kappa_{\rm p}=3$, $\beta=0$, (b) $\kappa_{\rm w}/\kappa_{\rm p}=3$, $\beta=1$, (c) $\kappa_{\rm w}/\kappa_{\rm p}=1$, $\beta=1$, (d) $\kappa_{\rm w}/\kappa_{\rm p}=0.1$, $\beta=1$.}}
    \label{fig:SI-temperature}
\end{figure}

\red{
The temperature fields given by Eqs.~\eqref{eq:SI-temp1} and \eqref{eq:SI-temp2} are characterized by the ratio of thermal conductivities $\kappa_{\rm w}/\kappa_{\rm p}$ and the degree of bias $\beta$. Figure~\ref{fig:SI-temperature} shows some typical cases: (a) $\kappa_{\rm w}/\kappa_{\rm p}=3$, $\beta=0$ (no bias), (b) $\kappa_{\rm w}/\kappa_{\rm p}=3$ (low thermal conductivity of the particle), $\beta=1$, (c) $\kappa_{\rm w}/\kappa_{\rm p}=1$ (equal thermal conductivities), $\beta=1$, (d) $\kappa_{\rm w}/\kappa_{\rm p}=0.1$, $\beta=1$ (high thermal conductivity of the particle). Panel (a) results in a radially symmetric profile, whereas panels (b--d) show the deformation of contours due to the bias $\beta\neq 0$. As the thermal conductivity of the light-absorbing particle increases (going from panel (b) to panel (d)), the temperature elevation inside the particle decreases, and the temperature distribution becomes more uniform, leading to a weaker temperature gradient. }

\end{document}